\documentclass{article}
\begin{document}
\begin{center}
{\LARGE\bf Nonstrange baryonia}

\large

\vskip3ex
S.M. Gerasyuta $ ^{*}$ and E.E. Matskevich $ ^{**}$

\vskip2ex
Department of Theoretical Physics, St. Petersburg State University,
198904,

St. Petersburg, Russia

and

Department of Physics, LTA, 194021, St. Petersburg, Russia

\vskip4ex

{\bf Abstract}
\end{center}
\vskip4ex
\large

The relativistic six-quark equations including the $u$, $d$ quarks and
antiquarks are found. The nonstrange baryonia $B \bar B$ are constructed
without the mixing of the quarks and antiquarks. The relativistic six-quark
amplitudes of the baryonia are calculated. The poles of these amplitudes
determine the masses of baryonia. 15 masses of baryonia are predicted. The
mass of baryonium with the spin-parity $J^P=0^-$ $M=1835\, MeV$ is used as
a fit.


\vskip2ex
\noindent
$ ^{*}$ gerasyuta@SG6488.spb.edu

\noindent
$ ^{**}$ matskev@pobox.spbu.ru

\vskip2ex
{\bf I. Introduction.}
\vskip2ex

BES Collaboration observed a significant threshold enhancement of $p\bar p$
mass spectrum in the radiative decay $J/\psi\to\gamma\, p\bar p$ [1].
Recently BES Collaboration reported the results on $X(1835)$
in the $J/\psi\to\gamma\, \eta'\pi^+ \pi^-$ channel [2]. Under the strong
assumption that the $p\bar p$ threshold enhancement and $X(1835)$ are the
same resonance, Zhu and Gao suggested $X(1835)$ could be a $J^{PC}=0^{-+}$
$I^G=0^+$ $p\bar p$ baryonium [3].

Theoretical investigations of baryon-antibaryon bound states date back
to the proposal of Fermi and Yang [4] to make the pion out of a
nucleon-antinucleon pair. The model of Nambu and Jona-Lasinio [5] which
is constructed to give a nearly zero-mass pion as a fermion-antifermion
bound state, also has a scalar resonance of twice the fermion mass.
Enhancement in the baryon-antibaryon channel near the threshold are expected
on the basis of duality arguments [6 -- 8] and by comparison with the
systematic of resonance formation in meson-meson and meson-baryon channels
[9]. A historical survey of bound states or resonances coupled to the
nucleon-antinucleon channel is given in Ref. [10]. Gluonic states can
couple to baryon-antibaryon channels of appropriate spin and parity.
The discussions of B decays involving baryon-antibaryon pairs include Refs.
[11 -- 15].

Theoretical work speculated many possibilities for the enhancement
such as the t-channel pion exchange, some kind of threshold kinematical
effects, as new resonance below threshold or $p\bar p$ bound state
[16 -- 23].

In a series of papers [24 -- 28] a method has been developed which is
convenient for analysing relativistic three-hadron systems. The physics of
the three-hadron system can be described by means of a pair interaction
between the particles. There are three isobar channels, each of which
consists of a two-particle isobar and the third particle. The presence
of the isobar representation together with the condition of unitarity in
the pair energies and of analyticity leads to a system of integral equations
in a single variable. Their solution makes it possible to describe the
interaction of the produced particles in three-hadron systems.

In Ref. [29] a representation of the Faddeev equation in the form of a
dispersion relation in the pair energy of the two interacting particles
was used. This was found to be convenient in order to obtain an approximate
solution of the Faddeev equation by a method based on extraction of the
leading singularities of the amplitude. With a rather crude approximation
of the low-energy $NN$ interaction a relatively good description of the
form factor of tritium (helium-3) at low $q^2$ was obtained.

In our papers [30 -- 32] relativistic generalization of the three-body
Faddeev equations was obtained in the form of dispersion relations in the
pair energy of two interacting quarks. The mass spectrum of $S$-wave
baryons including $u$, $d$, $s$ quarks was calculated by a method based on
isolating the leading singularities in the amplitude. We searched for the
approximate solution of integral three-quark equations by taking into
account two-particle and triangle singularities, all the weaker ones being
neglected. If we considered such approximation, which corresponds to
taking into account two-body and triangle singularities, and defined all
the smooth functions of the subenergy variables (as compared with the
singular part of the amplitude) in the middle point of the physical region
of Dalitz-plot, then the problem was reduced to the one of solving a system
of simple algebraic equations.

In the previous paper [35] the relativistic six-quark equations are found in
the framework of coupled-channel formalism. The dynamical mixing between
the subamplitudes of hexaquark are considered. The six-quark amplitudes
of dibaryons are calculated. The poles of these amplitudes determine the
masses of dibaryons. We calculated the contribution of six-quark
subamplitudes to the hexaquark amplitudes.

In the present paper the relativistic six-quark equations including $u$, $d$
quarks and antiquarks are found. The nonstrange barionia $B \bar B$ are
constructed without the mixing of the quarks and antiquarks. The
relativistic six-quark amplitudes of the baryonia are calculated.
The poles of these amplitudes determine the masses of baryonia. In Sec. II
we briefly discuss the relativistic Faddeev approach. The relativistic
three-quark equations are constructed in the form of the dispersion
relations over the two-body subenergy. The approximate solution of these
equations using the method based on the extraction of leading singularities
of the amplitude are obtained. We calculated the mass spectrum of $S$-wave
baryons with $J^P=\frac{1}{2}^+$, $\frac{3}{2}^+$ (Table I). In Sec. III the
six-quark amplitudes of baryonia are constructed. The dynamical mixing
between the subamplitudes of baryonia are considered. The relativistic
six-quark equations are obtained in the form of the dispersion relations
over the two-body subenergy. The approximate solutions of these equations
using the method based on the extraction of leading singularities of the
amplitude are obtained. Sec. IV is devoted to the calculation results for
the baryonia mass spectrum and the contributions of subamplitudes to the
baryonia amplitude (Tables II, III, IV). In conclusion, the status
of the considered model is discussed.

\vskip2ex
{\bf II. Brief introduction of relativistic Faddeev equations.}
\vskip2ex

We consider the derivation of the relativistic generalization of the
Faddeev equation for the example of the $\Delta$-isobar
($J^P=\frac{3}{2}^+$). This is convenient because the spin-flavour part
of the wave function of the $\Delta$-isobar contains only nonstrange quarks
and pair interactions with the quantum numbers of a $J^P=1^+$ diquark
(in the color state $\bar 3_c$). The $3q$ baryon state $\Delta$ is
constructed as color singlet. Suppose that there is a $\Delta$-isobar
current which produces three $u$ quarks (Fig. 1a). Successive pair
interactions lead to the diagrams shown in Fig. 1b-1f. These diagrams can
be grouped according to which of the three quark pairs undergoes the last
interaction i.e., the total amplitude can be represented as a sum of
diagrams. Taking into account the equality of all pair interactions of
nonstrange quarks in the state with $J^P=1^+$, we obtain the corresponding
equation for the amplitudes:

\begin{equation}
A_1 (s, s_{12}, s_{13}, s_{23})=\lambda+A_1 (s, s_{12})+
A_1 (s, s_{13})+A_1 (s, s_{23})\, . \end{equation}

\noindent
Here, the $s_{ik}$ are the pair energies of particles 1, 2 and 3, and $s$
is the total energy of the system. Using the diagrams of  Fig. 1, it is
easy to write down a graphical equation for the function $A_1 (s, s_{12})$
(Fig. 2). To write down a concrete equations for the function
$A_1 (s, s_{12})$ we must specify the amplitude of the pair interaction
of the quarks. We write the amplitude of the interaction of two quarks in
the state $J^P=1^+$ in the form:

\begin{equation}
a_1(s_{12})=\frac{G^2_1(s_{12})}
{1-B_1(s_{12})} \, ,\end{equation}

\begin{equation}
B_1(s_{12})=\int\limits_{4m^2}^{\infty}
\, \frac{ds'_{12}}{\pi}\frac{\rho_1(s'_{12})G^2_1(s'_{12})}
{s'_{12}-s_{12}} \, ,\end{equation}

\begin{eqnarray}
\rho_1 (s_{12})&=&
\left(\frac{1}{3}\, \frac{s_{12}}{4m^2}+\frac{1}{6}\right)
\left(\frac{s_{12}-4m^2}{s_{12}}\right)^{\frac{1}{2}} \, .
\end{eqnarray}

\noindent
Here $G_1(s_{12})$ is the vertex function of a diquark with $J^P=1^+$.
$B_1(s_{12})$ is the Chew-Mandelstam function [34], and $\rho_1 (s_{12})$
is the phase spaces for a diquark with $J^P=1^+$.

The pair quarks amplitudes $qq \to qq$ are calculated in the framework of
the dispersion $N/D$ method with the input four-fermion interaction
[35, 36] with the quantum numbers of the gluon [37, 38].

The four-quark interaction is considered as an input:

\begin{eqnarray}
 & g_V \left(\bar q \lambda I_f \gamma_{\mu} q \right)^2 +
2\, g^{(s)}_V \left(\bar q \lambda I_f \gamma_{\mu} q \right)
\left(\bar s \lambda \gamma_{\mu} s \right)+
g^{(ss)}_V \left(\bar s \lambda \gamma_{\mu} s \right)^2
 \, . & \end{eqnarray}

\noindent
Here $I_f$ is the unity matrix in the flavor space $(u, d)$. $\lambda$ are
the color Gell-Mann matrices. Dimensional constants of the four-fermion
interaction $g_V$, $g^{(s)}_V$ and $g^{(ss)}_V$ are parameters of the
model. At $g_V =g^{(s)}_V =g^{(ss)}_V$ the flavor $SU(3)_f$ symmetry occurs.
The strange quark violates the flavor $SU(3)_f$ symmetry. In order to
avoid additional violation parameters we introduce the scale of the
dimensional parameters [38]:

\begin{equation}
g=\frac{m^2}{\pi^2}g_V =\frac{(m+m_s)^2}{4\pi^2}g_V^{(s)} =
\frac{m_s^2}{\pi^2}g_V^{(ss)}
\, ,\end{equation}

\begin{eqnarray}
\Lambda=\frac{4\Lambda(ik)}
{(m_i+m_k)^2}. \nonumber \end{eqnarray}

\noindent
Here $m_i$ and $m_k$ are the quark masses in the intermediate state of
the quark loop. Dimensionless parameters $g$ and $\Lambda$ are supposed
to be constants which are independent of the quark interaction type. The
applicability of Eq. (5) is verified by the success of
De Rujula-Georgi-Glashow quark model [37], where only the short-range
part of Breit potential connected with the gluon exchange is responsible
for the mass splitting in hadron multiplets.

In the case under discussion the interacting pairs of particles do not
form bound states. Thefore, the integration in the dispersion integral (7)
run from $4m^2$ to $\infty$. The equation corresponding to Fig. 2 can be
written in the form:

\begin{eqnarray}
A_1(s,s_{12})&
=&\frac{\lambda_1 B_1(s_{12})}{1-B_1(s_{12})}+
\frac{G_1(s_{12})}{1-B_1(s_{12})}
\int\limits_{4m^2}^{\infty}
\, \frac{ds'_{12}}{\pi}\frac{\rho_1(s'_{12})}
{s'_{12}-s_{12}}G_1(s'_{12})\nonumber\\
&&\nonumber\\
 & \times & \int\limits_{-1}^{+1} \, \frac{dz}{2}
[A_1(s,s'_{13})+A_1(s,s'_{23})] \, .
\end{eqnarray}

In Eq. (7) $z$ is the cosine of the angle between the relative momentum
of particles 1 and 2 in the intermediate state and the momentum of the third
particle in the final state in the c.m.s. of the particles 1 and 2. In
our case of equal mass of the quarks 1, 2 and 3, $s'_{13}$ and $s'_{12}$
are related by the equation (8) (See Ref. [39])

\begin{eqnarray}
s'_{13}=2m^2-\frac{(s'_{12}+m^2-s)}{2}
\pm \frac{z}{2} \sqrt{\frac{(s'_{12}-4m^2)}{s'_{12}}
(s'_{12}-(\sqrt{s}+m)^2)(s'_{12}-(\sqrt{s}-m)^2)}\, .
\end{eqnarray}

The expression for $s'_{23}$ is similar to (8) with the replacement
$z\to -z$. This makes it possible to replace
$[A_1(s,s'_{13})+A_1(s,s'_{23})]$ in (7) by $2A_1(s,s'_{13})$.

From the amplitude $A_1(s,s_{12})$ we shall extract the singularities
of the diquark amplitude:

\begin{eqnarray}
A_1(s,s_{12})=
\frac{\alpha_1(s,s_{12}) B_1(s_{12})}{1-B_1(s_{12})} \, .
\end{eqnarray}

The equation for the reduced amplitude $\alpha_1(s,s_{12})$ can be written
as

\begin{eqnarray}
\alpha_1(s,s_{12})&=&\lambda+\frac{1}{B_1(s_{12})}
\int\limits_{4m^2}^{\infty}
\, \frac{ds'_{12}}{\pi}\frac{\rho_1(s'_{12})}
{s'_{12}-s_{12}}G_1(s'_{12})\int\limits_{-1}^{+1} \, \frac{dz}{2}
\, \frac{2\alpha_1(s,s'_{13}) B_1(s'_{13})}{1-B_1(s'_{13})} \, .
\end{eqnarray}

The next step is to include into (10) a cutoff at large $s'_{12}$. This
cutoff is needed to approximate the contribution of the interaction at
short distances. In this connection we shall rewrite Eq. (10) as

\begin{eqnarray}
\alpha_1(s,s_{12})&=&\lambda+\frac{1}{B_1(s_{12})}
\int\limits_{4m^2}^{\infty}
\, \frac{ds'_{12}}{\pi}\Theta(\Lambda-s'_{12})\frac{\rho_1(s'_{12})}
{s'_{12}-s_{12}}G_1\int\limits_{-1}^{+1} \, \frac{dz}{2}
\, \frac{2\alpha_1(s,s'_{13}) B_1(s'_{13})}{1-B_1(s'_{13})} \, .
\end{eqnarray}

In Eq. (11) we have chosen a hard cutoff. However, we can also use a soft
cutoff, for instance
$G_1(s'_{12})=G_1 \exp \left( -\frac{(s'_{12}-4m^2)^2}{\Lambda^2} \right)$,
which leaves the results of calculations of the mass spectrum essentially
unchanged.

The construction of the approximate solution of Eq. (11) is based on
extraction of the leading singularities are close to the region
$s_{ik}\approx 4m^2$. The structure of the singularities of amplitudes
with a different number of rescattering (Fig. 1) is the following [39].
The strongest singularities in $s_{ik}$ arise from pair rescatterings of
quarks: square-root singularity corresponding to a threshold and pole
singularities corresponding to bound states (on the first sheet in the case
of real bound states, and on the second sheet in the case of virtual bound
states). The diagrams of Figs. 1b and 1c have only these two-particle
singularities. In addition to two-particle singularities diagrams of
Figs. 1d and 1e have their own specific triangle singularities. The
diagram of Figs. 1f describes a larger number of three-particle
singularities. In addition to singularities of triangle type it contains
other weaker singularities. Such a classification of singularities
makes it possible to search for an approximate solution of Eq. (11),
taking into account a definite number of leading singularities and
neglecting the weaker ones. We use the approximation in which the
singularity corresponding to a single interaction of all three particles,
the triangle singularity, is taken into account.

For fixed values of $s$ and $s'_{12}$ the integration is carried out
over the region of the variable $s'_{13}$ corresponding to a physical
transition of the current into three quarks (the physical region of
Dalitz plot). It is convenient to take the central point of this region,
corresponding to $z=0$, to determinate the function $\alpha_1(s,s_{12})$
and also the Chew-Mandelstam function $B_1(s_{12})$ at the point
$s_{12}=s_0=\frac{s}{3}+m^2$. Then the equation for the $\Delta$ isobar
takes the form:

\begin{equation}
\alpha_1(s,s_0)=\lambda+I_{1,1}(s,s_0)\cdot 2\, \alpha_1(s,s_0)
\, , \end{equation}

\begin{equation}
I_{1,1}(s,s_0)=\int\limits_{4m^2}^{\Lambda_1}
\, \frac{ds'_{12}}{\pi} \frac{\rho_1(s'_{12})}
{s'_{12}-s_{12}}G_1\int\limits_{-1}^{+1} \, \frac{dz}{2}
\, \frac{G_1}{1-B_1(s'_{13})}
\, . \end{equation}

We can obtain an approximate solution of Eq. (14)

\begin{equation}
\alpha_1(s,s_0)=\lambda [1-2\, I_{1,1}(s,s_0)]^{-1}
\, . \end{equation}

The function $I_{1,1}(s,s_0)$ takes into account correctly the singularities
corresponding to the fact that all propagators of triangle diagrams like
those of Figs. 1d and 1e reduce to zero. The right-hand side of (14) may
have a pole in $s$, which corresponds to a bound state of the three quarks.
The choice of the cutoff $\Lambda$ makes it possible to fix the value of
the mass of the $\Delta$ isobar.

Baryons of $S$-wave multiplets have a completely symmetric spin-flavor
part of the wave function, and spin $\frac{3}{2}$ corresponds to the
decuplet which has a symmetric flavor part of the wave function. Octet
states have spin $\frac{1}{2}$ and a mixed symmetry of the flavor function.

In analogy with the case of the $\Delta$ isobar we can obtain the
rescattering amplitudes for all $S$-wave states with $J^P=\frac{3}{2}^+$,
which include quarks of various flavors. These amplitudes will satisfy
systems of integral equations. In considering the $J^P=\frac{1}{2}^+$
octet we must include the integration of the quarks in the $0^+$ and $1^+$
states (in the colour state $\bar 3_c$). Including all possible rescattering
of each pair of quarks and grouping the terms according to the final states
of the particles, we obtain the amplitudes $A_0$ and $A_1$, which satisfy
the corresponding systems of integral equations. If we choose the
approximation in which two-particle and triangle singularities are taken
into account, and if all functions which depend on the physical region of
the Dalitz plot, the problem of solving the system of integral equations
reduces to one of solving simple algebraic equations.

In our calculation the quark masses $m_u=m_d=m$ and $m_s$ are not uniquely
determined. In order to fix $m$ and $m_s$ anyhow, we make the simple
assumption that $m=\frac{1}{3} m_{\Delta} (1232)$
$m=\frac{1}{3} m_{\Omega} (1672)$. The strange quark breaks the flavor
$SU(3)_f$ symmetry (6).

In Ref. [32] we consider two versions of calculations. If the first version
the $SU(3)_f$ symmetry is broken by the scale shift of the dimensional
parameters. A single cutoff parameter in pair energy is introduced for all
diquark states $\lambda_1=12.2$.

In the Table I the calculated masses of the $S$-wave baryons are shown.
In the first version we use only three parameters: the subenergy cutoff
$\lambda$ and the vertex function $g_0$, $g_1$, which corresponds to the
quark-quark interaction in $0^+$ and $1^+$ states. In this case the mass
values of strange baryons with $J^P=\frac{1}{2}^+$ are less than the
experimental ones. This means that the contribution color-magnetic is too
large. In the second version we introduce four parameters: cutoff
$\lambda_0$, $\lambda_1$ and the vertex function $g_0$, $g_1$. We decrease
the color-magnetic interaction in $0^+$ strange channels and calculated mass
values of two baryonic multiplets  $J^P=\frac{1}{2}^+$, $\frac{3}{2}^+$
are in good agreement with the experimental data [40].

The essential difference between $\Sigma$ and $\Lambda$ is the spin of the
lighter diquark. The model explains both the sign and magnitude of this mass
splitting.

The suggested method of the approximate solution of the relativistic
three-quark equations allows us to calculate the $S$-wave baryons mass
spectrum. The interaction constants, determined the baryons spectrum in
our model, are similar to ones in the bootstrap quark model of $S$-wave
mesons [38]. The diquark interaction forces are defined by the gluon
exchange. The relative contribution of the instanton-induced interaction is
less than that with the gluon exchange. This is the consequence of
$1/N_c$-expansion [38].

The gluon exchange corresponds to the color-magnetic interaction, which
is responsible for the spin-spin splitting in the hadron models. The sign
of the color-magnetic term is such as to made any baryon of spin
$\frac{3}{2}$ heavier than its spin-$\frac{1}{2}$ counterpart (containing
the same flavors).

We manage with quarks as with real particles. However, in the soft region,
the quark diagrams should be treated as spectral integrals over quark masses
with the spectral density $\rho (m^2)$: the integration over quark masses
in the amplitudes puts away the quark singularities and introduced the
hadron ones. One can believe that the approximation:

\begin{equation}
\rho (m^2) \to \delta (m^2-m_q^2)
\end{equation}

\noindent
could be possible for the low-lying hadrons (here $m_q$ is the "mass" of
the constituent quark).

We hope the approach given by (15) is sufficiently good for the calculation
of the low-lying baryons being carried out here. The problem of
distribution over quark masses is important when one considers that the
high-excited states need spectral studies.

\vskip2ex
{\bf III. Six-quark amplitudes of the baryonia.}
\vskip2ex

We derive the relativistic six-quark equations in the framework of the
dispersion relation technique. We use only planar diagrams; the other
diagrams due to the rules of $1/N_c$ expansion [41 -- 43] are neglected.
The current generates a six-quark system. The correct equations for the
amplitude are obtained by taking into account all possible subamplitudes.
It corresponds to the division of complete system into subsystems with a
smaller number of particles. Then one should represent a six-particle
amplitude as a sum of 15 subamplitudes:

\begin{equation}
A=\sum\limits_{i<j \atop i, j=1}^6 A_{ij}\, . \end{equation}

This defines the division of the diagrams into groups according to the
certain pair interaction of particles. The total amplitude can be
represented graphically as a sum of diagrams. We need to consider only
one group of diagrams and the amplitude corresponding to them, for example
$A_{12}$. We shall consider the derivation of the relativistic
generalization of the Faddeev-Yakubovsky approach.

In our case the low-lying baryonia are considered. We take into account the
pairwise interaction of all quarks and antiquarks in the baryonia.

For instance, we consider the $1^{uu}$-diquarks with spin-parity
$J^P=1^+$ for the baryonium content $uuu\bar d \bar d \bar d$ (Fig. 3). The
set of diagrams associated with the amplitude $A_{12}$ can further be broken
down into five groups corresponding to subamplitudes:
$A_1^{1^{uu}}(s,s_{12345},s_{1234},s_{123},s_{12})$,
$A_1^{1^{\bar d \bar d}}(s,s_{12345},s_{1234},s_{123},s_{12})$,
$A_1^{1^{u \bar d}}(s,s_{12345},s_{1234},s_{123},s_{12})$, baryonium
$A_2^{1^{uu}1^{\bar d \bar d}}(s,s_{12345},s_{1234},s_{12},s_{34})$,
$A_3^{1^{uu}1^{u\bar d}1^{\bar d \bar d}}
(s,s_{12345},s_{12},s_{34},s_{56})$.
Here $s_{ik}$ is the two-particle subenergy squared, $s_{ijk}$ corresponds
to the energy squared of particles $i$, $j$, $k$, $s_{ijkl}$ is the energy
squared of particles $i$, $j$, $k$, $l$, $s_{ijklm}$ corresponds to the
energy squared of particles $i$, $j$, $k$, $l$, $m$ and $s$ is the system
total energy squared.

The amplitude $A_1^{1^{uu}}(s,s_{12345},s_{1234},s_{123},s_{12})$ consists
of the five color sub-structures: the diquark $1^{uu}$ in the color state
$\bar 3_c$, the quark $u$ in the color state $3_c$, the three $\bar d$
antiquarks each in color state $\bar 3_c$, therefore we obtain
$\bar 3_c \times 3_c=1_c +8_c$,
$\bar 3_c \times \bar 3_c \times \bar 3_c=1_c +8_c +8_c +10_c^*$.
Then we consider the total color singlet.

The baryonium amplitude
$A_2^{1^{uu}1^{\bar d \bar d}}(s,s_{12345},s_{1234},s_{12},s_{34})$
contains the diquark in the color state $\bar 3_c$, antidiquark
$1^{\bar d \bar d}$ in the color state $3_c$, $u$-quark in the color state
$3_c$, $\bar d$-antiquark in color state $\bar 3_c$. We use the following
equations: $\bar 3_c \times 3_c=1_c +8_c$, $3_c \times \bar 3_c=1_c +8_c$.
Then the baryonium amplitude is the total singlet.

The amplitude
$A_3^{1^{uu}1^{u\bar d}1^{\bar d \bar d}}(s,s_{12345},s_{12},s_{34},s_{56})$
consists of the diquark $1^{uu}$ in the color state $\bar 3_c$, antidiquark
$1^{\bar d \bar d}$ in color state $3_c$ and noncolor state $1^{u\bar d}$,
therefore the total color singlet can be constructed.

The subamplitudes
$A_1^{1^{\bar d \bar d}}(s,s_{12345},s_{1234},s_{123},s_{12})$ and
$A_1^{1^{u \bar d}}(s,s_{12345},s_{1234},s_{123},s_{12})$ are also the
color singlets.

The system of graphical equations Fig. 3 is determined using the
selfconsistent method. The coefficients are determined by the permutation
of quarks [44, 45]. We should discuss the coefficient multiplying of the
diagrams in the equations of Fig. 3. For example, we consider the first
subamplitude $A_1^{1^{uu}}(s,s_{12345},s_{1234},s_{123},s_{12})$. In the
Fig. 3 the first coefficient is equal to 2 (permutation particles 1 and 2).
The second coefficient is equal to $6=2$ (permutation particles 1 and 2)
$\times 3$ (we consider the third, the fifth, the sixth particles). The
similar approach allows us to take into account the coefficients in all
equations.

In order to represent the subamplitudes $A_1^{1^{uu}}$, $A_1^{1^{u\bar d}}$,
$A_1^{1^{\bar d \bar d}}$, $A_2^{1^{uu}1^{\bar d \bar d}}$,
$A_3^{1^{uu}1^{u\bar d}1^{\bar d \bar d}}$ in the form of dispersion
relations, it is necessary to define the amplitudes of $qq$ and $q\bar q$
interactions. This is similar to the three quark case (Sec. II).

We use the results of our relativistic quark model [38] and write down the
pair quark amplitudes in the form:

\begin{equation}
a_n(s_{ik})=\frac{G^2_n(s_{ik})}
{1-B_n(s_{ik})} \, ,\end{equation}

\begin{equation}
B_n(s_{ik})=\int\limits_{(m_i+m_k)^2}^{\frac{(m_i+m_k)^2\Lambda}{4}}
\hskip2mm \frac{ds'_{ik}}{\pi}\frac{\rho_n(s'_{ik})G^2_n(s'_{ik})}
{s'_{ik}-s_{ik}} \, .\end{equation}

\noindent
Here $G_n(s_{ik})$ are the diquark vertex functions (Table V). The vertex
functions are determined by the contribution of the crossing channels.
The vertex functions satisfy the Fierz relations. These vertex
functions are generated from $g_V$, $g^{(s)}_V$ and $g^{(ss)}_V$.
$B_n(s_{ik})$ and $\rho_n (s_{ik})$ are the Chew-Mandelstam functions with
cutoff $\Lambda$ [34] and the phase spaces, respectively:

\begin{eqnarray}
\rho_n (s_{ik},J^{PC})&=&\left(\alpha(n,J^{PC}) \frac{s_{ik}}{(m_i+m_k)^2}
+\beta(n,J^{PC})+\delta(n,J^{PC}) \frac{(m_i-m_k)^2}{s_{ik}}\right)
\nonumber\\
&&\nonumber\\
&\times & \frac{\sqrt{(s_{ik}-(m_i+m_k)^2)(s_{ik}-(m_i-m_k)^2)}}
{s_{ik}}\, .
\end{eqnarray}

The coefficients $\alpha(n,J^{PC})$, $\beta(n,J^{PC})$ and
$\delta(n,J^{PC})$ are given in Table V.

Here $n=1$ coresponds to $q\bar q$-pairs with $J^P=0^-$, $n=2$ corresponds
to the $q\bar q$-pairs with $J^P=1^-$, $n=3$ defines the $qq$-pairs with
$J^P=0^+$, $n=4$ coresponds to $J^P=1^+$ $qq$ states.

In the case in question the interacting quarks do not produce a bound
state, therefore the integration in (20) -- (24) is carried out from
the threshold $(m_i+m_k)^2$ to the cutoff $\Lambda(ik)$.

The coupled integral equations correspond to Fig. 3 can be described as:

\begin{eqnarray}
A_1^{1^{uu}}(s,s_{12345},s_{1234},s_{123},s_{12})&
=&\frac{\lambda_1 B_{1^{uu}}(s_{12})}{[1-B_{1^{uu}}(s_{12})]}+
2\hat J_1(s_{12},1^{uu}) A_1^{1^{uu}}(s,s_{12345},s_{1234},s_{123},s'_{13})
\nonumber\\
&&\nonumber\\
&+&6\hat J_1(s_{12},1^{uu})
A_1^{1^{u \bar d}}(s,s_{12345},s_{1234},s_{123},s'_{13}) \, ,
\\
&&\nonumber\\
A_1^{1^{\bar d \bar d}}(s,s_{12345},s_{1234},s_{123},s_{12})&
=&\frac{\lambda_1 B_{1^{\bar d \bar d}}(s_{12})}
{[1-B_{1^{\bar d \bar d}}(s_{12})]}+
2\hat J_1(s_{12},1^{\bar d \bar d})
A_1^{1^{\bar d \bar d}}(s,s_{12345},s_{1234},s_{123},s'_{13})
\nonumber\\
&&\nonumber\\
&+&6\hat J_1(s_{12},1^{\bar d \bar d})
A_1^{1^{u \bar d}}(s,s_{12345},s_{1234},s_{123},s'_{13}) \, ,
\\
&&\nonumber\\
A_1^{1^{u \bar d}}(s,s_{12345},s_{1234},s_{123},s_{12})&
=&\frac{\lambda_1 B_{1^{u \bar d}}(s_{12})}{[1-B_{1^{u \bar d}}(s_{12})]}+
2\hat J_1(s_{12},1^{u \bar d})
A_1^{1^{uu}}(s,s_{12345},s_{1234},s_{123},s'_{13})
\nonumber\\
&&\nonumber\\
&+&2\hat J_1(s_{12},1^{u \bar d})
A_1^{1^{\bar d \bar d}}(s,s_{12345},s_{1234},s_{123},s'_{13})
\nonumber\\
&&\nonumber\\
&+&4\hat J_1(s_{12},1^{u \bar d})
A_1^{1^{u \bar d}}(s,s_{12345},s_{1234},s_{123},s'_{13})
\nonumber\\
&&\nonumber\\
&+&4\hat J_2(s_{12},1^{u \bar d})
A_2^{1^{uu}1^{\bar d \bar d}}(s,s_{12345},s_{1234},s'_{13},s'_{24}) \, ,
\\
&&\nonumber\\
A_2^{1^{uu}1^{\bar d \bar d}}(s,s_{12345},s_{1234},s_{12},s_{34})&
=&\frac{\lambda_2 B_{1^{uu}}(s_{12})
B_{1^{\bar d \bar d}}(s_{34})}
{[1-B_{1^{uu}}(s_{12})][1-B_{1^{\bar d \bar d}}(s_{34})]}
\nonumber\\
&&\nonumber\\
&+&2\hat J_4(s_{12},s_{34},1^{uu},1^{\bar d \bar d})
A_1^{1^{uu}}(s,s_{12345},s_{1235},s_{125},s'_{15})
\nonumber\\
&&\nonumber\\
&+&2\hat J_4(s_{12},s_{34},1^{\bar d \bar d},1^{uu})
A_1^{1^{\bar d \bar d}}(s,s_{12345},s_{1235},s_{125},s'_{15})
\nonumber\\
&&\nonumber\\
&+&4\hat J_3(s_{12},s_{34},1^{uu},1^{\bar d \bar d})
A_1^{1^{u \bar d}}(s,s_{12345},s_{1234},s'_{123},s'_{23})
\nonumber\\
&&\nonumber\\
&+&4\hat J_6(s_{12},s_{34},1^{uu},1^{uu})
A_2^{1^{uu}1^{\bar d \bar d}}(s,s_{12456},s_{1456},s'_{15},s'_{46})
\nonumber\\
&&\nonumber\\
&+&4\hat J_8(s_{12},s_{34},1^{uu},1^{\bar d \bar d})
A_3^{1^{uu}1^{u \bar d}1^{\bar d \bar d}}
(s,s_{12345},s'_{15},s'_{23},s'_{46})
 \, ,\\
&&\nonumber\\
A_3^{1^{uu}1^{u \bar d}1^{\bar d \bar d}}(s,s_{12345},s_{12},s_{34},s_{56})&
=&\frac{\lambda_3 B_{1^{uu}}(s_{12})
B_{1^{u \bar d}}(s_{34}) B_{1^{\bar d \bar d}}(s_{56})}
{[1- B_{1^{uu}}(s_{12})]
[1- B_{1^{u \bar d}}(s_{34})] [1- B_{1^{\bar d \bar d}}(s_{56})]}
\nonumber \\
&&\nonumber\\
&+&2\hat J_9(s_{12},s_{34},s_{56},1^{uu},1^{u \bar d},1^{\bar d \bar d})
A_1^{1^{uu}}(s,s_{12345},s_{1234},s'_{123},s'_{23})
\nonumber \\
&&\nonumber\\
&+&2\hat J_9(s_{12},s_{34},s_{56},1^{\bar d \bar d},1^{u \bar d},1^{uu})
A_1^{1^{\bar d \bar d}}(s,s_{12345},s_{1234},s'_{123},s'_{23})
\nonumber\\
&&\nonumber\\
&+&2\hat J_9(s_{12},s_{34},s_{56},1^{uu},1^{u \bar d},1^{\bar d \bar d})
A_1^{1^{u \bar d}}(s,s_{12345},s_{1234},s'_{123},s'_{23})
\nonumber \\
&&\nonumber\\
&+&4\hat J_9(s_{12},s_{34},s_{56},1^{uu},1^{\bar d \bar d},1^{u \bar d})
A_1^{1^{u \bar d}}(s,s_{12345},s_{1234},s'_{123},s'_{23})
\nonumber \\
&&\nonumber\\
&+&2\hat J_9(s_{12},s_{34},s_{56},1^{\bar d \bar d},1^{u \bar d},1^{uu})
A_1^{1^{u \bar d}}(s,s_{12345},s_{1234},s'_{123},s'_{23})
\nonumber \\
&&\nonumber\\
&+&4\hat J_{10}(s_{12},s_{34},s_{56},1^{uu},1^{u \bar d},1^{\bar d \bar d})
A_2^{1^{uu}1^{\bar d \bar d}}(s,s_{12345},s_{2345},s'_{23},s'_{45})
\, ,\nonumber \\
&&
\end{eqnarray}

\noindent
where

\begin{eqnarray}
\hat J_1(s_{12},i)&=&\frac{G_i(s_{12})}{[1- B_i(s_{12})]}
\int\limits_{(m_1+m_2)^2}^{\frac{(m_1+m_2)^2\Lambda_i}{4}}
\frac{ds'_{12}}{\pi}\frac{G_i(s'_{12})\rho_i(s'_{12})}
{s'_{12}-s_{12}} \int\limits_{-1}^{+1} \frac{dz_1(1)}{2} \, ,\\
&&\nonumber\\
\hat J_2(s_{12},i)&=&\frac{G_i(s_{12})}{[1- B_i(s_{12})]}
\int\limits_{(m_1+m_2)^2}^{\frac{(m_1+m_2)^2\Lambda_i}{4}}
\frac{ds'_{12}}{\pi}\frac{G_i(s'_{12})\rho_i(s'_{12})}
{s'_{12}-s_{12}}
\frac{1}{2\pi}\int\limits_{-1}^{+1}\frac{dz_1(2)}{2}
\int\limits_{-1}^{+1} \frac{dz_2(2)}{2}
\nonumber\\
&&\nonumber\\
&\times&
\int\limits_{z_3(2)^-}^{z_3(2)^+} dz_3(2)
\frac{1}{\sqrt{1-z_1^2(2)-z_2^2(2)-z_3^2(2)+2z_1(2) z_2(2) z_3(2)}} \, , \\
&&\nonumber\\
\hat J_3(s_{12},s_{34},i,j)&=&\frac{G_i(s_{12})G_j(s_{34})}
{[1- B_i(s_{12})][1- B_j(s_{34})]}
\int\limits_{(m_1+m_2)^2}^{\frac{(m_1+m_2)^2\Lambda_i}{4}}
\frac{ds'_{12}}{\pi}\frac{G_i(s'_{12})\rho_i(s'_{12})}
{s'_{12}-s_{12}}\nonumber\\
&&\nonumber\\
&\times&\int\limits_{(m_3+m_4)^2}^{\frac{(m_3+m_4)^2\Lambda_j}{4}}
\frac{ds'_{34}}{\pi}\frac{G_j(s'_{34})\rho_j(s'_{34})}
{s'_{34}-s_{34}}
\int\limits_{-1}^{+1} \frac{dz_1(3)}{2} \int\limits_{-1}^{+1}
\frac{dz_2(3)}{2} \, , \\
&&\nonumber\\
\hat J_4(s_{12},s_{34},i,j)&=&\frac{B_j(s_{34})}{[1- B_j(s_{34})]}
\hat J_1(s_{12},i) \, , \\
&&\nonumber\\
\hat J_6(s_{12},s_{34},i,j)&=&\hat J_1(s_{12},i) \cdot \hat J_1(s_{34},j)
 \, , \\
&&\nonumber\\
\hat J_8(s_{12},s_{34},i,j)&=&\frac{G_i(s_{12})G_j(s_{34})}
{[1- B_i(s_{12})][1- B_j(s_{34})]}
\int\limits_{(m_1+m_2)^2}^{\frac{(m_1+m_2)^2\Lambda_i}{4}}
\frac{ds'_{12}}{\pi}\frac{G_i(s'_{12})\rho_i(s'_{12})}
{s'_{12}-s_{12}}\nonumber\\
&&\nonumber\\
&\times&\int\limits_{(m_3+m_4)^2}^{\frac{(m_3+m_4)^2\Lambda_j}{4}}
\frac{ds'_{34}}{\pi}\frac{G_j(s'_{34})\rho_j(s'_{34})}
{s'_{34}-s_{34}}\nonumber\\
&&\nonumber\\
&\times&\frac{1}{(2\pi)^2}\int\limits_{-1}^{+1}\frac{dz_1(8)}{2}
\int\limits_{-1}^{+1} \frac{dz_2(8)}{2}
\int\limits_{-1}^{+1} \frac{dz_3(8)}{2}
\int\limits_{z_4(8)^-}^{z_4(8)^+} dz_4(8)
\int\limits_{-1}^{+1} \frac{dz_5(8)}{2}
\int\limits_{z_6(8)^-}^{z_6(8)^+} dz_6(8)
\nonumber\\
&&\nonumber\\
&\times&
\frac{1}{\sqrt{1-z_1^2(8)-z_3^2(8)-z_4^2(8)+2z_1(8) z_3(8) z_4(8)}}
\nonumber\\
&&\nonumber\\
&\times&
\frac{1}{\sqrt{1-z_2^2(8)-z_5^2(8)-z_6^2(8)+2z_2(8) z_5(8) z_6(8)}}
 \, , \\
&&\nonumber\\
\hat J_9(s_{12},s_{34},s_{56},i,j,k)&=&\frac{B_k(s_{56})}{[1- B_k(s_{56})]}
\hat J_3(s_{12},s_{34},i,j) \, , \\
&&\nonumber\\
\hat J_{10}(s_{12},s_{34},s_{56},i,j,k)&=
&\frac{G_i(s_{12})G_j(s_{34})G_k(s_{56})}
{[1- B_i(s_{12})][1- B_j(s_{34})][1- B_k(s_{56})]}
\int\limits_{(m_1+m_2)^2}^{\frac{(m_1+m_2)^2\Lambda_i}{4}}
\frac{ds'_{12}}{\pi}\frac{G_i(s'_{12})\rho_i(s'_{12})}{s'_{12}-s_{12}}
 \nonumber\\
&&\nonumber\\
&\times&
\int\limits_{(m_3+m_4)^2}^{\frac{(m_3+m_4)^2\Lambda_j}{4}}
\frac{ds'_{34}}{\pi}\frac{G_j(s'_{34})\rho_j(s'_{34})}
{s'_{34}-s_{34}}\int\limits_{(m_5+m_6)^2}^{\frac{(m_5+m_6)^2\Lambda_k}{4}}
\frac{ds'_{56}}{\pi}\frac{G_k(s'_{56})\rho_k(s'_{56})}{s'_{56}-s_{56}}
\nonumber\\
&&\nonumber\\
&\times&
\frac{1}{2\pi}\int\limits_{-1}^{+1}\frac{dz_1(10)}{2}
\int\limits_{-1}^{+1} \frac{dz_2(10)}{2}
\int\limits_{-1}^{+1} \frac{dz_3(10)}{2}
\int\limits_{-1}^{+1} \frac{dz_4(10)}{2}
\int\limits_{z_5(1-)^-}^{z_5(10)^+} dz_5(10)
\nonumber\\
&&\nonumber\\
&\times&
\frac{1}{\sqrt{1-z_1^2(10)-z_4^2(10)-z_5^2(10)+2z_1(10) z_4(10) z_5(10)}}
 \, .
\end{eqnarray}

In the equation (25) $z_1(1)$ is the cosine of the angle between the
relative momentum of particles 1 and 2 in the intermediate state and the
momentum of the particle 3 in the final state taken in the c.m. of particles
1 and 2. We can go from the integration of the cosine of the angle
$dz_1(1)$ to the integration over the subenergy $ds'_{13}$.

In Eq. (26) $z_1(2)$ is the cosine of the angle between the
relative momentum of particles 1 and 2 in the intermediate state and the
momentum of the particle 3 in the final state taken in the c.m. of particles
1 and 2, $z_2(2)$ is the cosine of the angle between the momenta of
particles 3 and 4 in the final state of c.m. of particles 1 and 2,
$z_3(2)$ is cosine of the angle between the relative momentum of particles
1 and 2 in the intermediate state and the momentum of the particle 4 in the
final state of c.m. of particles 1 and 2. Then we pass from
$dz_1(2)dz_2(2)dz_3(2)$ to $ds'_{13}ds'_{34}ds'_{24}$.

In Eq. (27) $z_1(3)$ is the cosine of the angle between the relative
momentum of particles 1, 2 in the intermediate state and the relative
momentum of particles 3, 4 in the intermediate state in c.m. of particles
3 and 4; $z_2(3)$ is the cosine of the angle between momentum of particle 3
in the intermediate state and relative momentum of particles 1, 2 in the
intermediate state in c.m. 1 and 2. We pass from $dz_1(3)dz_2(3)$ to
$ds'_{123}ds'_{23}$. The similar method is used for the functions
(28), (29), (31).

In Eq. (30) $z_1(8)$ is the cosine of the angle between momentum of particle
5 in the final state and the relative momentum of particles 1, 2 in the
intermediate state in c.m. of particles 1 and 2; $z_2(8)$ is the cosine
of the angle between the relative momentum of particles 1, 2 in the
intermediate state and the relative momentum of particles 3, 4 in the
intermediate state in c.m. of particles 3 and 4; $z_3(8)$ is the cosine of
the angle between momentum of particle 3 in the intermediate state and the
momentum of particle 5 in the final state in c.m. of particles 1 and 2;
$z_4(8)$ is the cosine of the angle between the momentum of particle 3 in
the intermediate state and the relative momentum of particles 1, 2 in the
intermediate state in c.m. of particles 1 and 2; $z_5(8)$ is the cosine
of angle between momentum of particle 6 in the final state and the relative
momentum of particles 1, 2 in the intermediate state in c.m. of particles
3 and 4; $z_6(8)$ is the cosine of the angle between momentum of particle
6 in the final state and the relative momentum of particles 3, 4 in the
intermediate state in c.m. of particles 3 and 4. We pass from
$dz_1(8)dz_2(8)dz_3(8)dz_4(8)dz_5(8)dz_6(8)$ to
$ds'_{15}ds'_{123}ds'_{35}ds'_{23}ds'_{126}ds'_{46}$.

In Eq. (32) $z_1(10)$ is the cosine of angle between relative momentum of
particles 1, 2 in the intermediate state and the relative momentum of
particles 3, 4 in the intermediate state in c.m. of particles 3 and 4;
$z_2(10)$ is the cosine of angle between the relative momentum of particles
1, 2 in the intermediate state and momentum of particle 3 in the final state
in c.m. of particles 1 and 2; $z_3(10)$ is the cosine of the angle between
the relative momentum of the particles 3, 4 in the intermediate state and
the relative momentum of particles 5, 6 in the intermediate state in c.m.
of particles 5 and 6; $z_4(10)$ is the cosine of angle between relative
momentum of particles 1, 2 in the intermediate state and the momentum of
particle 5 in the final state in c.m. of particles 3 and 4; $z_5(10)$ is
the cosine of the angle between the relative momentum of the particles 3, 4
in the intermediate state and the momentum of particle 5 in the final state
in c.m. of particles 3 and 4. We pass from
$dz_1(10)dz_2(10)dz_3(10)dz_4(10)dz_5(10)$ to
$ds'_{123}ds'_{23}ds'_{345}ds'_{125}ds'_{45}$.

Let us extract two- and three-particle singularities in the amplitudes
$A_1^{1^{uu}}(s,s_{12345},s_{1234},s_{123},s_{12})$,
$A_1^{1^{\bar d \bar d}}(s,s_{12345},s_{1234},s_{123},s_{12})$,
$A_1^{1^{u \bar d}}(s,s_{12345},s_{1234},s_{123},s_{12})$,
$A_2^{1^{uu}1^{\bar d \bar d}}(s,s_{12345},s_{1234},s_{12},s_{34})$,\\
$A_3^{1^{uu}1^{u \bar d}1^{\bar d \bar d}}
(s,s_{12345},s_{12},s_{34},s_{56})$:

\begin{eqnarray}
A_1^{1^{uu}}(s,s_{12345},s_{1234},s_{123},s_{12})&=
&\frac{\alpha_1^{1^{uu}} (s,s_{12345},s_{1234},s_{123},s_{12})
B_{1^{uu}}(s_{12})}{[1-B_{1^{uu}}(s_{12})]} \, ,\\
&&\nonumber\\
A_1^{1^{\bar d \bar d}}(s,s_{12345},s_{1234},s_{123},s_{12})&=
&\frac{\alpha_1^{1^{\bar d \bar d}} (s,s_{12345},s_{1234},s_{123},s_{12})
B_{1^{\bar d \bar d}}(s_{12})}{[1-B_{1^{\bar d \bar d}}(s_{12})]} \, ,\\
&&\nonumber\\
A_1^{1^{u \bar d}}(s,s_{12345},s_{1234},s_{123},s_{12})&=
&\frac{\alpha_1^{1^{u \bar d}} (s,s_{12345},s_{1234},s_{123},s_{12})
B_{1^{u \bar d}}(s_{12})}{[1-B_{1^{u \bar d}}(s_{12})]} \, ,\\
&&\nonumber\\
A_2^{1^{uu}1^{\bar d \bar d}}(s,s_{12345},s_{1234},s_{12},s_{34})&=
&\frac{\alpha_2^{1^{uu}1^{\bar d \bar d}}
(s,s_{12345},s_{1234},s_{12},s_{34})
B_{1^{uu}}(s_{12})B_{1^{\bar d \bar d}}(s_{34})}{[1-B_{1^{uu}}(s_{12})]
[1-B_{1^{\bar d \bar d}}(s_{34})]} \, , \\
&&\nonumber\\
A_3^{1^{uu}1^{u \bar d}1^{\bar d \bar d}}(s,s_{12345},s_{12},s_{34},s_{56})
&=&\frac{\alpha_3^{1^{uu}1^{u \bar d}1^{\bar d \bar d}}
(s,s_{12345},s_{12},s_{34},s_{56})
B_{1^{uu}}(s_{12})B_{1^{u \bar d}}(s_{34}) B_{1^{\bar d \bar d}}(s_{56})}
{[1- B_{1^{uu}}(s_{12})] [1- B_{1^{u \bar d}}(s_{34})]
[1- B_{1^{\bar d \bar d}}(s_{56})]}
 \, . \nonumber\\
&&
\end{eqnarray}

We do not extract four-particles singularities, because they are weaker
than two- and three-particle singularities.

We used the classification of singularities, which was proposed in
paper [39]. The construction of the approximate solution of Eqs.
(33) -- (37) is based on the extraction of the leading singularities
of the amplitudes. The main singularities in $s_{ik}=(m_i+m_k)^2$
are from pair rescattering of the particles $i$ and $k$. First of all there
are threshold square-root singularities. Also possible are pole
singularities which correspond to the bound states. The diagrams of Fig. 3
apart from two-particle singularities have triangular singularities and the
singularities defining the interactions of four, five and six particles.
Such classification allows us to search the corresponding solution of Eqs.
(20) -- (24) by taking into account some definite number of leading
singularities and neglecting all the weaker ones. We consider the
approximation which defines two-particle, triangle and four-,  five- and
six-particle singularities. The contribution of two-particle and triangle
singularities are more important, but we must take into account also the
other singularities.

The five functions $\alpha_i$ are the smooth functions of $s_{ik}$,
$s_{ijk}$, $s_{ijkl}$ $s_{ijklm}$ as compared with the singular part of the
amplitudes, hence they can be expanded in a series in the singularity point
and only the first term of this series should be employed further. Using
this classification, one defines the reduced amplitudes $\alpha_i$ as well
as the $B$-functions in the middle point of physical region of Dalitz-plot
at the point $s_0$:

\begin{eqnarray}
s_0=\frac{s+4\sum\limits_{i=1}^{6} m_i^2}
{\sum\limits_{i,k=1 \atop i<k}^{6} m_{ik}^2}
\, ,
\end{eqnarray}

\begin{eqnarray}
s_{123}=s_0 \sum\limits_{i,k=1 \atop i<k}^{3} m_{ik}^2
-\sum\limits_{i=1}^{3} m_i^2
\, ,
\end{eqnarray}

\begin{eqnarray}
s_{1234}=s_0 \sum\limits_{i,k=1 \atop i<k}^{4} m_{ik}^2
-2\sum\limits_{i=1}^{4} m_i^2
\, .
\end{eqnarray}

Such choice of point $s_0$ allows us to replace integral equations
(20) -- (24) (Fig. 3) by the algebraic equations (41) -- (45), respectively:

\begin{eqnarray}
\alpha_1^{1^{uu}} &=&\lambda+2 I_1(1^{uu}1^{uu}) \alpha_1^{1^{uu}}
+6 I_1(1^{uu}1^{u \bar d}) \alpha_1^{1^{u \bar d}}
\, , \\
&&\nonumber\\
\alpha_1^{1^{\bar d \bar d}} &=&\lambda
+2 I_1(1^{\bar d \bar d}1^{\bar d \bar d}) \alpha_1^{1^{\bar d \bar d}}
+6 I_1(1^{\bar d \bar d}1^{u \bar d}) \alpha_1^{1^{u \bar d}}
\, , \\
&&\nonumber\\
\alpha_1^{1^{u \bar d}} &=&\lambda
+2 I_1(1^{u \bar d}1^{uu}) \alpha_1^{1^{uu}}
+2 I_1(1^{u \bar d}1^{\bar d \bar d}) \alpha_1^{1^{\bar d \bar d}}
+4 I_1(1^{u \bar d}1^{u \bar d}) \alpha_1^{1^{u \bar d}}
+4 I_2(1^{u \bar d}1^{uu}1^{\bar d \bar d})
\alpha_2^{1^{uu}1^{\bar d \bar d}}
\, , \nonumber\\
&&\\
\alpha_2^{1^{uu}1^{\bar d \bar d}} &=&\lambda
+2 I_4(1^{uu}1^{\bar d \bar d}1^{uu}) \alpha_1^{1^{uu}}
+2 I_4(1^{\bar d \bar d}1^{uu}1^{\bar d \bar d})
\alpha_1^{1^{\bar d \bar d}}
+4 I_3(1^{uu}1^{\bar d \bar d}1^{u\bar d}) \alpha_1^{1^{u \bar d}}
\nonumber\\
&&\nonumber\\
&+&4 I_6(1^{uu}1^{\bar d \bar d}1^{uu}1^{\bar d \bar d})
\alpha_2^{1^{uu}1^{\bar d \bar d}}
+4 I_8(1^{uu}1^{\bar d \bar d}1^{uu}1^{u \bar d}1^{\bar d \bar d})
\alpha_3^{1^{uu}1^{u \bar d}1^{\bar d \bar d}}
\, , \nonumber\\
&& \\
\alpha_3^{1^{uu}1^{u \bar d}1^{\bar d \bar d}}&=&\lambda
+2 I_9(1^{uu}1^{u \bar d}1^{\bar d \bar d}1^{uu}) \alpha_1^{1^{uu}}
+2 I_9(1^{\bar d \bar d}1^{u \bar d}1^{uu}1^{\bar d \bar d})
\alpha_1^{1^{\bar d \bar d}}
+2 I_9(1^{uu}1^{u \bar d}1^{\bar d \bar d}1^{u \bar d})
\alpha_1^{1^{u \bar d}}
\nonumber\\
&&\nonumber\\
&+&4 I_9(1^{uu}1^{\bar d \bar d}1^{u \bar d}1^{u \bar d})
\alpha_1^{1^{u \bar d}}
+2 I_9(1^{\bar d \bar d}1^{u \bar d}1^{uu}1^{u \bar d})
\alpha_1^{1^{u \bar d}}
+4 I_{10}(1^{uu}1^{u \bar d}1^{\bar d \bar d}1^{uu}1^{\bar d \bar d})
\alpha_2^{1^{uu}1^{\bar d \bar d}} \, , \nonumber\\
&&
\end{eqnarray}

\noindent
where $\lambda_i$ are the current constants. We used the functions
$I_1$, $I_2$, $I_3$, $I_4$, $I_6$, $I_8$, $I_9$, $I_{10}$:

\begin{eqnarray}
I_1(ij)&=&\frac{B_j(s_0^{13})}{B_i(s_0^{12})}
\int\limits_{(m_1+m_2)^2}^{\frac{(m_1+m_2)^2\Lambda_i}{4}}
\frac{ds'_{12}}{\pi}\frac{G_i^2(s_0^{12})\rho_i(s'_{12})}
{s'_{12}-s_0^{12}} \int\limits_{-1}^{+1} \frac{dz_1(1)}{2}
\frac{1}{1-B_j (s'_{13})}\, , \\
&&\nonumber\\
I_2(ijk)&=&\frac{B_j(s_0^{13}) B_k(s_0^{24})}{B_i(s_0^{12})}
\int\limits_{(m_1+m_2)^2}^{\frac{(m_1+m_2)^2\Lambda_i}{4}}
\frac{ds'_{12}}{\pi}\frac{G_i^2(s_0^{12})\rho_i(s'_{12})}
{s'_{12}-s_0^{12}}
\frac{1}{2\pi}\int\limits_{-1}^{+1}\frac{dz_1(2)}{2}
\int\limits_{-1}^{+1} \frac{dz_2(2)}{2}\nonumber\\
&&\nonumber\\
&\times&
\int\limits_{z_3(2)^-}^{z_3(2)^+} dz_3(2)
\frac{1}{\sqrt{1-z_1^2(2)-z_2^2(2)-z_3^2(2)+2z_1(2) z_2(2) z_3(2)}}
\nonumber\\
&&\nonumber\\
&\times& \frac{1}{1-B_j (s'_{13})} \frac{1}{1-B_k (s'_{24})}
 \, , \\
&&\nonumber\\
I_3(ijk)&=&\frac{B_k(s_0^{23})}{B_i(s_0^{12}) B_j(s_0^{34})}
\int\limits_{(m_1+m_2)^2}^{\frac{(m_1+m_2)^2\Lambda_i}{4}}
\frac{ds'_{12}}{\pi}\frac{G_i^2(s_0^{12})\rho_i(s'_{12})}
{s'_{12}-s_0^{12}}\nonumber\\
&&\nonumber\\
&\times&\int\limits_{(m_3+m_4)^2}^{\frac{(m_3+m_4)^2\Lambda_j}{4}}
\frac{ds'_{34}}{\pi}\frac{G_j^2(s_0^{34})\rho_j(s'_{34})}
{s'_{34}-s_0^{34}}
\int\limits_{-1}^{+1} \frac{dz_1(3)}{2} \int\limits_{-1}^{+1}
\frac{dz_2(3)}{2} \frac{1}{1-B_k (s'_{23})} \, , \\
&&\nonumber\\
I_4(ijk)&=&I_1(ik) \, , \\
&&\nonumber\\
I_6(ijkl)&=&I_1(ik) \cdot I_1(jl)
 \, , \\
&&\nonumber\\
I_8(ijklm)&=&\frac{B_k(s_0^{15})B_l(s_0^{23})B_m(s_0^{46})}
{B_i(s_0^{12}) B_j(s_0^{34})}
\int\limits_{(m_1+m_2)^2}^{\frac{(m_1+m_2)^2\Lambda_i}{4}}
\frac{ds'_{12}}{\pi}\frac{G_i^2(s_0^{12})\rho_i(s'_{12})}
{s'_{12}-s_0^{12}}\nonumber\\
&&\nonumber\\
&\times&\int\limits_{(m_3+m_4)^2}^{\frac{(m_3+m_4)^2\Lambda_j}{4}}
\frac{ds'_{34}}{\pi}\frac{G_j^2(s_0^{34})\rho_j(s'_{34})}
{s'_{34}-s_0^{34}}\nonumber\\
&&\nonumber\\
&\times&\frac{1}{(2\pi)^2}\int\limits_{-1}^{+1}\frac{dz_1(8)}{2}
\int\limits_{-1}^{+1} \frac{dz_2(8)}{2}
\int\limits_{-1}^{+1} \frac{dz_3(8)}{2}
\int\limits_{z_4(8)^-}^{z_4(8)^+} dz_4(8)
\int\limits_{-1}^{+1} \frac{dz_5(8)}{2}
\int\limits_{z_6(8)^-}^{z_6(8)^+} dz_6(8)
\nonumber\\
&&\nonumber\\
&\times&
\frac{1}{\sqrt{1-z_1^2(8)-z_3^2(8)-z_4^2(8)+2z_1(8) z_3(8) z_4(8)}}
\nonumber\\
&&\nonumber\\
&\times&
\frac{1}{\sqrt{1-z_2^2(8)-z_5^2(8)-z_6^2(8)+2z_2(8) z_5(8) z_6(8)}}
\nonumber\\
&&\nonumber\\
&\times& \frac{1}{1-B_k (s'_{15})} \frac{1}{1-B_l (s'_{23})}
\frac{1}{1-B_m (s'_{46})}
 \, , \\
&&\nonumber\\
I_9(ijkl)&=&I_3(ijl) \, , \\
&&\nonumber\\
I_{10}(ijklm)&=
&\frac{B_l(s_0^{23})B_m(s_0^{45})}
{B_i(s_0^{12}) B_j(s_0^{34}) B_k(s_0^{56})}
\int\limits_{(m_1+m_2)^2}^{\frac{(m_1+m_2)^2\Lambda_i}{4}}
\frac{ds'_{12}}{\pi}\frac{G_i^2(s_0^{12})\rho_i(s'_{12})}{s'_{12}-s_0^{12}}
\nonumber\\
&&\nonumber\\
&\times&
\int\limits_{(m_3+m_4)^2}^{\frac{(m_3+m_4)^2\Lambda_j}{4}}
\frac{ds'_{34}}{\pi}\frac{G_j^2(s_0^{34})\rho_j(s'_{34})}
{s'_{34}-s_0^{34}}
\int\limits_{(m_5+m_6)^2}^{\frac{(m_5+m_6)^2\Lambda_k}{4}}
\frac{ds'_{56}}{\pi}\frac{G_k^2(s_0^{56})\rho_k(s'_{56})}{s'_{56}-s_0^{56}}
\nonumber\\
&&\nonumber\\
&\times&
\frac{1}{2\pi}\int\limits_{-1}^{+1}\frac{dz_1(10)}{2}
\int\limits_{-1}^{+1} \frac{dz_2(10)}{2}
\int\limits_{-1}^{+1} \frac{dz_3(10)}{2}
\int\limits_{-1}^{+1} \frac{dz_4(10)}{2}
\int\limits_{z_5(1-)^-}^{z_5(10)^+} dz_5(10)
\nonumber\\
&&\nonumber\\
&\times&
\frac{1}{\sqrt{1-z_1^2(10)-z_4^2(10)-z_5^2(10)+2z_1(10) z_4(10) z_5(10)}}
\nonumber\\
&&\nonumber\\
&\times& \frac{1}{1-B_l (s'_{23})} \frac{1}{1-B_m (s'_{45})}
 \, ,
\end{eqnarray}

\noindent
where $i$, $j$, $k$, $l$, $m$ correspond to the diquarks with the
spin-parity $J^P=0^+, 1^+$ and mesons with the spin-parity
$J^P=0^-, 1^-$.

The other choices of point $s_0$ do not change essentially the contributions
of $\alpha_i$, therefore we omit the indices $s_0^{ik}$. Since the
vertex functions depend only slightly on energy, it is possible to treat
them as constants in our approximation.

The solutions of the system of equations are considered as:

\begin{equation}
\alpha_i(s)=F_i(s,\lambda_i)/D(s) \, ,\end{equation}

\noindent
where zeros of $D(s)$ determinants define the masses of bound states of
baryonia.

As example, we consider the equations for the quark content
$uuu \bar d \bar d \bar d$ with the isospin $I=3$ and the spin-parity
$J^P=3^-$ (Fig. 3). The similar equations have been calculated for the
isospin $I=0,\, 1, \, 2,\, 3$ and the spin-parity $J^P=0^-, 1^-, 2^-, 3^-$.
We take into account the $u$ and $d$ quarks.

In Appendix I the reduced amplitudes of baryonium $uud \bar u \bar u \bar d$
$IJ=00$ are given.

\vskip2ex
{\bf IV. Calculation results.}
\vskip2ex

The poles of the five reduced amplitudes $\alpha_i$ correspond to the
bound state and determine the mass of the baryonium with the quark content
$uuu \bar d \bar d \bar d$, with the isospin $I=3$ and the spin-parity
$J^P=3^-$. The quark mass of model $m=410\, MeV$ coincides with the ordinary
baryon one in our model (Sec. II).

The model in question has only two parameters: the cutoff parameter
$\Lambda=11$ (similar to the three quark model (Sec. II)) and the gluon
coupling constant $g=0.314$. This parameter is determined by the baryonium
mass $M=1835\, MeV$. The estimation of theoretical error on the
baryonia masses is $1\, MeV$. This result was obtained by the choice of
model parameters.

We predict the degeneracy of the some states. In the Table II the calculated
masses of nonstrange baryonia are shown. The contributions of subamplitudes
to the six-quark amplitude are shown in the Appendix I (for example, the
baryonium with the mass $M=1835\, MeV$). The states
($\Delta \bar \Delta+\Delta \bar n+n \bar \Delta+n\bar n$) and
($\Delta \bar \Delta+\Delta \bar p+p \bar \Delta+p\bar p$) with the isospin
$I=0$ and the spin-parity $J^P=0^-$ possess the mass $M=1835\, MeV$.
For the ($\Delta \bar \Delta+n \bar \Delta$) and
($\Delta \bar \Delta+\Delta \bar n$) with the isospin $I=1$ and the
spin-parity $J^P=0^-$ we obtained the mass $M=1928\, MeV$.

We predict the degeneracy of baryonia
$M(uud \bar d \bar d \bar d,\, I=2)=M(uuu \bar u \bar d \bar d,\, I=2)
=M(udd \bar d \bar d \bar d,\, I=1)=M(uuu \bar u \bar u \bar d,\, I=1)$.
For the states
$M(uud \bar u \bar d \bar d,\, I=1)=M(udd \bar u \bar d \bar d,\, I=0)
=M(uud \bar u \bar u \bar d,\, I=0)$ and
$M(uuu \bar u \bar u \bar u,\, I=0)=M(ddd \bar d \bar d \bar d,\, I=0)$
the degeneracy is also obtained.

\vskip2ex
{\bf V. Conclusion.}
\vskip2ex

A somewhat simple picture of baryonium is that of a deuteron-like $N\bar N$
bound state or resonance, benefiting from the attractive potential mediated
by the exchange of gluon [33]. We do not consider the influence of
annihilation on the spectrum.

Entem and Fernandez, describing scattering data and mass shifts of $p\bar p$
levels in a constituent quark model, assign the threshold enhancement to
final-state interaction [46, 47]. Zou and Chiang find that final state
interaction makes an important contribution to the $p\bar p$ near threshold
enhancement [48].

The baryonium state with $M=1835\, MeV$ is considered as $p\bar p$ state [3]
or the second radial excitation of $\eta'$ meson [49].

In our case this state have following content
$\Delta \bar \Delta+\Delta \bar p+p \bar \Delta+p\bar p$ with isospin
$I=0$ and spin-parity $J^P=0^-$.

We calculated the masses of baryonia with the isospin $I=0,\, 1,\, 2,\, 3$
and spin-parity $J^P=0^-,\, 1^-,\, 2^-,\, 3^-$ (Table II).

The interesting reseach is the contribution of baryonia consisting of
$u,d,s$-quarks and antiquarks.

\vskip2.0ex
{\bf Acknowledgments.}
\vskip2.0ex

The authors would like to thank T. Barnes and C.-Y. Wong for useful
discussions. The work was carried with the support of the Russian Ministry
of Education (grant 2.1.1.68.26).

\newpage

{\bf \Large References.}

\vskip5ex

\noindent
1. J.Z. Bai et al, BES Collaboration, Phys. Rev. Lett. {\bf 91}, 022001
(2003).

\noindent
2. M. Ablikim et al, BES Collaboration, Phys. Rev. Lett. {\bf 95}, 262001
(2005).

\noindent
3. S.L. Zhu and C.S. Gao, Commun. Theor. Phys. {\bf 46}, 291 (2006).

\noindent
4. E. Fermi and C.N. Yang, Phys. Rev. {\bf 76}, 1739 (1949).

\noindent
5. Y. Nambu and G. Jona-Lasinio, Phys. Rev. {\bf 122}, 345 (1961);
Phys. Rev. {\bf 124}, 246 (1961).

\noindent
6. J.L. Rosner, Phys. Rev. Lett. {\bf 21}, 950 (1968).

\noindent
7. H. Harari, Phys. Rev. Lett. {\bf 22}, 562 (1969).

\noindent
8. J.L. Rosner, Phys. Rev. Lett. {\bf 22}, 689 (1969).

\noindent
9. J.L. Rosner, Phys. Rev. D{\bf 6}, 2717 (1972).

\noindent
10. J.-M. Richard, Nucl. Phys. B Proc. Suppl. {\bf 86}, 361 (2000).

\noindent
11. I. Dunietz, Phys. Rev. D{\bf 58}, 094010 (1998).

\noindent
12. W.-S. Hou and A. Soni, Phys. Rev. Lett. {\bf 86}, 4247 (2001).

\noindent
13. C.-K. Chua, W.-S. Hou and S.-Y. Tsai, Phys. Rev. D{\bf 65}, 034003
(2002); Phys. Lett. B{\bf 528},

233 (2002); Phys. Rev. D{\bf 66}, 054004
(2002).

\noindent
14. G.J. Ding and M.L. Yan, Phys. Rev. C{\bf 72}, 015208 (2005).

\noindent
15. B. Loiseau and S. Wycech, Phys. Rev. C{\bf 72}, 011001 (2005).

\noindent
16. B. Loiseau and S. Wycech, Int. J. Mod. Phys. A{\bf 20}, 1990 (2005).

\noindent
17. C.H. Chang and H.R. Pang, Commun. Theor. Phys. {\bf 43}, 275 (2005).

\noindent
18. X.G. He, X.Q. Li and J.P. Ma, Phys. Rev. D{\bf 71}, 014031 (2005).

\noindent
19. D.V. Bugg, Phys. Lett. B{\bf 598}, 8 (2004).

\noindent
20. I.N. Mishustin, L.M. Satarov, T.J. Burvenich, H. Stoecker and
W. Greiner, Phys. Rev. C{\bf 71},

035201 (2005).

\noindent
21. B. Kerbikov, A. Stavinsky and V. Fedotov, Phys. Rev. C{\bf 69}, 055205
(2004).

\noindent
22. A. Datta and P.J. O'Donnel, Phys. Lett. B{\bf 567}, 273 (2003).

\noindent
23. J.L. Rosner, Phys. Rev. D{\bf 68}, 014004 (2003).

\noindent
24. I.J.R. Aitchison, J. Phys. G{\bf 3}, 121 (1977).

\noindent
25. J.J. Brehm, Ann. Phys. (N.Y.) {\bf 108}, 454 (1977).

\noindent
26. I.J.R. Aitchison and J.J. Brehm, Phys. Rev. D{\bf 17}, 3072 (1978).

\noindent
27. I.J.R. Aitchison and J.J. Brehm,
Phys. Rev. D{\bf 20}, 1119, 1131 (1979).

\noindent
28. J.J. Brehm, Phys. Rev. D{\bf 21}, 718 (1980).

\noindent
29. A.V. Anisovich and V.V. Anisovich, Yad. Fiz. {\bf 53}, 1485 (1991)
[Sov. J. Nucl. Phys. {\bf 53}, 915

(1991)].

\noindent
30. S.M. Gerasyuta, Yad. Fiz. {\bf 55}, 3030 (1992) [Sov. J. Nucl. Phys.
{\bf 55}, 1693 (1992)].

\noindent
31. S.M. Gerasyuta, Nuovo Cimento Soc. Ital. Fis. A{\bf 106}, 37 (1993).

\noindent
32. S.M. Gerasyuta, Z. Phys. C{\bf 60}, 683 (1993).

\noindent
33. S.M. Gerasyuta and E.E. Matskevich, Phys. Rev. D{\bf 82}, 056002 (2010).

\noindent
34. G. Chew, S. Mandelstam, Phys. Rev. {\bf 119}, 467 (1960).

\noindent
35. T. Appelqvist and J.D. Bjorken, Phys. Rev. D{\bf 4}, 3726 (1971).

\noindent
36. C.C. Chiang, C.B. Chiu, E.C.G. Sudarshan and X. Tata,
Phys. Rev. D{\bf 25}, 1136 (1982).

\noindent
37. A. De Rujula, H. Georgi and S.L. Glashow, Phys. Rev. D{\bf 12}, 147
(1975).

\noindent
38. V.V. Anisovich, S.M. Gerasyuta and A.V. Sarantsev,
Int. J. Mod. Phys. A{\bf 6}, 625 (1991).

\noindent
39. V.V. Anisovich and A.A. Anselm, Usp. Fiz. Nauk {\bf 88}, 287 (1966)
[Sov. Phys. Usp. {\bf 9}, 117

(1966)].

\noindent
40. K. Nakamura el al. (Particle Data Group), J. Phys. G{\bf 37}, 075021
(2010).

\noindent
41. G.'t Hooft, Nucl. Phys. B{\bf 72}, 461 (1974).

\noindent
42. G. Veneziano, Nucl. Phys. B{\bf 117}, 519 (1976).

\noindent
43. E. Witten, Nucl. Phys. B{\bf 160}, 57 (1979).

\noindent
44. O.A. Yakubovsky, Sov. J. Nucl. Phys. {\bf 5}, 1312 (1967).

\noindent
45. S.P. Merkuriev and L.D. Faddeev, Quantum Scattering Theory for System
of Few Particles

(Nauka, Moscow 1985) p. 398.

\noindent
46. D.R. Entem and F. Fernandez, Eur. Phys. J. A{\bf 31}, 649 (2007).

\noindent
47. D.R. Entem and F. Fernandez, Phys. Rev. D{\bf 75}, 014004 (2007).

\noindent
48. B.S. Zou and H.C. Chiang, Phys. Rev. D{\bf 69}, 034004 (2004).

\noindent
49. T. Huang and S.-L. Zhu, Phys. Rev. D{\bf 73}, 014023 (2006).

\newpage

{\bf Appendix I. The reduced amplitudes of baryonium $IJ=00(1835)$.}

\vskip2.0ex

\begin{eqnarray}
\alpha_1^{1^{uu}}&=&\lambda
+2\, \alpha_1^{1^{u \bar u}} I_1(1^{uu}1^{u \bar u})
+2\, \alpha_1^{1^{u \bar d}} I_1(1^{uu}1^{u \bar d})
+2\, \alpha_1^{0^{ud}} I_1(1^{uu}0^{ud})
+4\, \alpha_1^{0^{u \bar u}} I_1(1^{uu}0^{u \bar u})
\nonumber\\
&&\nonumber\\
&+&2\, \alpha_1^{0^{u \bar d}} I_1(1^{uu}0^{u \bar d})
\nonumber\\
&&\nonumber\\
\alpha_1^{1^{\bar u \bar u}}&=&\lambda
+2\, \alpha_1^{1^{u \bar u}} I_1(1^{\bar u \bar u}1^{u \bar u})
+2\, \alpha_1^{1^{d \bar u}} I_1(1^{\bar u \bar u}1^{d \bar u})
+2\, \alpha_1^{0^{\bar u \bar d}} I_1(1^{\bar u \bar u}0^{\bar u \bar d})
+4\, \alpha_1^{0^{u \bar u}} I_1(1^{\bar u \bar u}0^{u \bar u})
\nonumber\\
&&\nonumber\\
&+&2\, \alpha_1^{0^{d \bar u}} I_1(1^{\bar u \bar u}0^{d \bar u})
\nonumber\\
&&\nonumber\\
\alpha_1^{1^{u \bar u}}&=&\lambda
+\alpha_1^{1^{uu}} I_1(1^{u \bar u}1^{uu})
+\alpha_1^{1^{\bar u \bar u}} I_1(1^{u \bar u}1^{\bar u \bar u})
+2\, \alpha_1^{1^{u \bar u}} I_1(1^{u \bar u}1^{u \bar u})
+\alpha_1^{1^{u \bar d}} I_1(1^{u \bar u}1^{u \bar d})
+\alpha_1^{1^{d \bar u}} I_1(1^{u \bar u}1^{d \bar u})
\nonumber\\
&&\nonumber\\
&+&\alpha_1^{0^{ud}} I_1(1^{u \bar u}0^{ud})
+\alpha_1^{0^{\bar u \bar d}} I_1(1^{u \bar u}0^{\bar u \bar d})
+2\, \alpha_1^{0^{u \bar u}} I_1(1^{u \bar u}0^{u \bar u})
+\alpha_1^{0^{u \bar d}} I_1(1^{u \bar u}0^{u \bar d})
+\alpha_1^{0^{d \bar u}} I_1(1^{u \bar u}0^{d \bar u})
\nonumber\\
&&\nonumber\\
&+&\alpha_2^{1^{uu}0^{\bar u \bar d}}
I_2(1^{u \bar u}1^{uu}0^{\bar u \bar d})
+\alpha_2^{1^{\bar u \bar u}0^{ud}}
I_2(1^{u \bar u}1^{\bar u \bar u}0^{ud})
+\alpha_2^{0^{ud}0^{\bar u \bar d}}
I_2(1^{u \bar u}0^{ud}0^{\bar u \bar d})
\nonumber\\
&&\nonumber\\
\alpha_1^{1^{u \bar d}}&=&\lambda
+\alpha_1^{1^{uu}} I_1(1^{u \bar d}1^{uu})
+\alpha_1^{1^{u \bar u}} I_1(1^{u \bar d}1^{u \bar u})
+\alpha_1^{1^{u \bar d}} I_1(1^{u \bar d}1^{u \bar d})
+\alpha_1^{1^{d \bar d}} I_1(1^{u \bar d}1^{d \bar d})
+\alpha_1^{0^{ud}} I_1(1^{u \bar d}0^{ud})
\nonumber\\
&&\nonumber\\
&+&2\, \alpha_1^{0^{\bar u \bar d}} I_1(1^{u \bar d}0^{\bar u \bar d})
+2\, \alpha_1^{0^{u \bar u}} I_1(1^{u \bar d}0^{u \bar u})
+\alpha_1^{0^{u \bar d}} I_1(1^{u \bar d}0^{u \bar d})
+\alpha_1^{0^{d \bar d}} I_1(1^{u \bar d}0^{d \bar d})
\nonumber\\
&&\nonumber\\
&+&2\, \alpha_2^{1^{uu}0^{\bar u \bar d}}
I_2(1^{u \bar d}1^{uu}0^{\bar u \bar d})
+2\, \alpha_2^{0^{ud}0^{\bar u \bar d}}
I_2(1^{u \bar d}0^{ud}0^{\bar u \bar d})
\nonumber\\
&&\nonumber\\
\alpha_1^{1^{d \bar u}}&=&\lambda
+\alpha_1^{1^{\bar u \bar u}} I_1(1^{d \bar u}1^{\bar u \bar u})
+\alpha_1^{1^{u \bar u}} I_1(1^{d \bar u}1^{u \bar u})
+\alpha_1^{1^{d \bar u}} I_1(1^{d \bar u}1^{d \bar u})
+\alpha_1^{1^{d \bar d}} I_1(1^{d \bar u}1^{d \bar d})
+2\, \alpha_1^{0^{ud}} I_1(1^{d \bar u}0^{ud})
\nonumber\\
&&\nonumber\\
&+&\alpha_1^{0^{\bar u \bar d}} I_1(1^{d \bar u}0^{\bar u \bar d})
+2\, \alpha_1^{0^{u \bar u}} I_1(1^{d \bar u}0^{u \bar u})
+\alpha_1^{0^{d \bar u}} I_1(1^{d \bar u}0^{d \bar u})
+\alpha_1^{0^{d \bar d}} I_1(1^{d \bar u}0^{d \bar d})
\nonumber\\
&&\nonumber\\
&+&2\, \alpha_2^{1^{\bar u \bar u}0^{ud}}
I_2(1^{d \bar u}1^{\bar u \bar u}0^{ud})
+2\, \alpha_2^{0^{ud}0^{\bar u \bar d}}
I_2(1^{d \bar u}0^{ud}0^{\bar u \bar d})
\nonumber\\
&&\nonumber\\
\alpha_1^{1^{d \bar d}}&=&\lambda
+\alpha_1^{1^{u \bar d}} I_1(1^{d \bar d}1^{u \bar d})
+\alpha_1^{1^{d \bar u}} I_1(1^{d \bar d}1^{d \bar u})
+2\, \alpha_1^{0^{ud}} I_1(1^{d \bar d}0^{ud})
+2\, \alpha_1^{0^{\bar u \bar d}} I_1(1^{d \bar d}0^{\bar u \bar d})
\nonumber\\
&&\nonumber\\
&+&2\, \alpha_1^{0^{u \bar d}} I_1(1^{d \bar d}0^{u \bar d})
+2\, \alpha_1^{0^{d \bar u}} I_1(1^{d \bar d}0^{d \bar u})
+2\, \alpha_2^{0^{ud}0^{\bar u \bar d}}
I_2(1^{d \bar d}0^{ud}0^{\bar u \bar d})
\nonumber\\
&&\nonumber\\
\alpha_1^{0^{ud}}&=&\lambda
+\alpha_1^{1^{uu}} I_1(0^{ud}1^{uu})
+2\, \alpha_1^{1^{u \bar u}} I_1(0^{ud}1^{u \bar u})
+\alpha_1^{1^{u \bar d}} I_1(0^{ud}1^{u \bar d})
+2\, \alpha_1^{1^{d \bar u}} I_1(0^{ud}1^{d \bar u})
+\alpha_1^{1^{d \bar d}} I_1(0^{ud}1^{d \bar d})
\nonumber\\
&&\nonumber\\
&+&\alpha_1^{0^{ud}} I_1(0^{ud}0^{ud})
+2\, \alpha_1^{0^{u \bar u}} I_1(0^{ud}0^{u \bar u})
+\alpha_1^{0^{u \bar d}} I_1(0^{ud}0^{u \bar d})
+2\, \alpha_1^{0^{d \bar u}} I_1(0^{ud}0^{d \bar u})
+\alpha_1^{0^{d \bar d}} I_1(0^{ud}0^{d \bar d})
\nonumber\\
&&\nonumber\\
\alpha_1^{0^{\bar u \bar d}}&=&\lambda
+\alpha_1^{1^{\bar u \bar u}} I_1(0^{\bar u \bar d}1^{\bar u \bar u})
+2\, \alpha_1^{1^{u \bar u}} I_1(0^{\bar u \bar d}1^{u \bar u})
+2\, \alpha_1^{1^{u \bar d}} I_1(0^{\bar u \bar d}1^{u \bar d})
+\alpha_1^{1^{d \bar u}} I_1(0^{\bar u \bar d}1^{d \bar u})
+\alpha_1^{1^{d \bar d}} I_1(0^{\bar u \bar d}1^{d \bar d})
\nonumber\\
&&\nonumber\\
&+&\alpha_1^{0^{\bar u \bar d}} I_1(0^{\bar u \bar d}0^{\bar u \bar d})
+2\, \alpha_1^{0^{u \bar u}} I_1(0^{\bar u \bar d}0^{u \bar u})
+2\, \alpha_1^{0^{u \bar d}} I_1(0^{\bar u \bar d}0^{u \bar d})
+\alpha_1^{0^{d \bar u}} I_1(0^{\bar u \bar d}0^{d \bar u})
+\alpha_1^{0^{d \bar d}} I_1(0^{\bar u \bar d}0^{d \bar d})
\nonumber\\
&&\nonumber\\
\alpha_1^{0^{u \bar u}}&=&\lambda
+\alpha_1^{1^{uu}} I_1(0^{u \bar u}1^{uu})
+\alpha_1^{1^{\bar u \bar u}} I_1(0^{u \bar u}1^{\bar u \bar u})
+2\, \alpha_1^{1^{u \bar u}} I_1(0^{u \bar u}1^{u \bar u})
+\alpha_1^{1^{u \bar d}} I_1(0^{u \bar u}1^{u \bar d})
+\alpha_1^{1^{d \bar u}} I_1(0^{u \bar u}1^{d \bar u})
\nonumber\\
&&\nonumber\\
&+&\alpha_1^{0^{ud}} I_1(0^{u \bar u}0^{ud})
+\alpha_1^{0^{\bar u \bar d}} I_1(0^{u \bar u}0^{\bar u \bar d})
+2\, \alpha_1^{0^{u \bar u}} I_1(0^{u \bar u}0^{u \bar u})
+\alpha_1^{0^{u \bar d}} I_1(0^{u \bar u}0^{u \bar d})
+\alpha_1^{0^{d \bar u}} I_1(0^{u \bar u}0^{d \bar u})
\nonumber\\
&&\nonumber\\
&+&\alpha_2^{1^{uu}1^{\bar u \bar u}}
I_2(0^{u \bar u}1^{uu}1^{\bar u \bar u})
+\alpha_2^{1^{uu}0^{\bar u \bar d}}
I_2(0^{u \bar u}1^{uu}0^{\bar u \bar d})
+\alpha_2^{1^{\bar u \bar u}0^{ud}}
I_2(0^{u \bar u}1^{\bar u \bar u}0^{ud})
+\alpha_2^{0^{ud}0^{\bar u \bar d}}
I_2(0^{u \bar u}0^{ud}0^{\bar u \bar d})
\nonumber\\
&&\nonumber\\
\alpha_1^{0^{u \bar d}}&=&\lambda
+\alpha_1^{1^{uu}} I_1(0^{u \bar d}1^{uu})
+2\, \alpha_1^{1^{u \bar u}} I_1(0^{u \bar d}1^{u \bar u})
+\alpha_1^{1^{u \bar d}} I_1(0^{u \bar d}1^{u \bar d})
+\alpha_1^{1^{d \bar d}} I_1(0^{u \bar d}1^{d \bar d})
+\alpha_1^{0^{ud}} I_1(0^{u \bar d}0^{ud})
\nonumber\\
&&\nonumber\\
&+&2\, \alpha_1^{0^{\bar u \bar d}} I_1(0^{u \bar d}0^{\bar u \bar d})
+2\, \alpha_1^{0^{u \bar u}} I_1(0^{u \bar d}0^{u \bar u})
+\alpha_1^{0^{u \bar d}} I_1(0^{u \bar d}0^{u \bar d})
+\alpha_1^{0^{d \bar d}} I_1(0^{u \bar d}0^{d \bar d})
\nonumber\\
&&\nonumber\\
&+&\alpha_2^{1^{uu}0^{\bar u \bar d}}
I_2(0^{u \bar d}1^{uu}0^{\bar u \bar d})
+2\, \alpha_2^{0^{ud}0^{\bar u \bar d}}
I_2(0^{u \bar d}0^{ud}0^{\bar u \bar d})
\nonumber\\
&&\nonumber\\
\alpha_1^{0^{d \bar u}}&=&\lambda
+\alpha_1^{1^{\bar u \bar u}} I_1(0^{d \bar u}1^{\bar u \bar u})
+2\, \alpha_1^{1^{u \bar u}} I_1(0^{d \bar u}1^{u \bar u})
+\alpha_1^{1^{d \bar u}} I_1(0^{d \bar u}1^{d \bar u})
+\alpha_1^{1^{d \bar d}} I_1(0^{d \bar u}1^{d \bar d})
+2\, \alpha_1^{0^{ud}} I_1(0^{d \bar u}0^{ud})
\nonumber\\
&&\nonumber\\
&+&\alpha_1^{0^{\bar u \bar d}} I_1(0^{d \bar u}0^{\bar u \bar d})
+2\, \alpha_1^{0^{u \bar u}} I_1(0^{d \bar u}0^{u \bar u})
+\alpha_1^{0^{d \bar u}} I_1(0^{d \bar u}0^{d \bar u})
+\alpha_1^{0^{d \bar d}} I_1(0^{d \bar u}0^{d \bar d})
+\alpha_2^{1^{\bar u \bar u}0^{ud}}
I_2(0^{d \bar u}1^{\bar u \bar u}0^{ud})
\nonumber\\
&&\nonumber\\
&+&2\, \alpha_2^{0^{ud}0^{\bar u \bar d}}
I_2(0^{d \bar u}0^{ud}0^{\bar u \bar d})
\nonumber\\
&&\nonumber\\
\alpha_1^{0^{d \bar d}}&=&\lambda
+2\, \alpha_1^{1^{u \bar d}} I_1(0^{d \bar d}1^{u \bar d})
+2\, \alpha_1^{1^{d \bar u}} I_1(0^{d \bar d}1^{d \bar u})
+2\, \alpha_1^{0^{ud}} I_1(0^{d \bar d}0^{ud})
+2\, \alpha_1^{0^{\bar u \bar d}} I_1(0^{d \bar d}0^{\bar u \bar d})
\nonumber\\
&&\nonumber\\
&+&2\, \alpha_1^{0^{u \bar d}} I_1(0^{d \bar d}0^{u \bar d})
+2\, \alpha_1^{0^{d \bar u}} I_1(0^{d \bar d}0^{d \bar u})
+4\, \alpha_2^{0^{ud}0^{\bar u \bar d}}
I_2(0^{d \bar d}0^{ud}0^{\bar u \bar d})
\nonumber\\
&&\nonumber\\
\alpha_2^{1^{uu}1^{\bar u \bar u}}&=&\lambda
+2\, \alpha_1^{0^{ud}} I_4(1^{uu}1^{\bar u \bar u}0^{ud})
+2\, \alpha_1^{0^{\bar u \bar d}}
I_4(1^{\bar u \bar u}1^{uu}0^{\bar u \bar d})
+4\, \alpha_1^{0^{u \bar u}} I_3(1^{uu}1^{\bar u \bar u}0^{u \bar u})
\nonumber\\
&&\nonumber\\
&+&4\, \alpha_2^{0^{ud}0^{\bar u \bar d}}
I_6(1^{uu}1^{\bar u \bar u}0^{ud}0^{\bar u \bar d})
+4\, \alpha_3^{0^{ud}0^{u \bar u}0^{\bar u \bar d}}
I_8(1^{uu}1^{\bar u \bar u}0^{ud}0^{u \bar u}0^{\bar u \bar d})
\nonumber\\
&&\nonumber\\
\alpha_2^{1^{uu}0^{\bar u \bar d}}&=&\lambda
+\alpha_1^{1^{\bar u \bar u}} I_4(0^{\bar u \bar d}1^{uu}1^{\bar u \bar u})
+2\, \alpha_1^{1^{u \bar u}} I_3(1^{uu}0^{\bar u \bar d}1^{u \bar u})
+2\, \alpha_1^{1^{u \bar d}} I_3(1^{uu}0^{\bar u \bar d}1^{u \bar d})
+2\, \alpha_1^{0^{ud}} I_4(1^{uu}0^{\bar u \bar d}0^{ud})
\nonumber\\
&&\nonumber\\
&+&\alpha_1^{0^{\bar u \bar d}}I_4(0^{\bar u \bar d}1^{uu}0^{\bar u \bar d})
+2\, \alpha_1^{0^{u \bar u}} I_3(1^{uu}0^{\bar u \bar d}0^{u \bar u})
+2\, \alpha_1^{0^{u \bar d}} I_3(1^{uu}0^{\bar u \bar d}0^{u \bar d})
+2\, \alpha_2^{1^{\bar u \bar u}0^{ud}}
I_6(1^{uu}0^{\bar u \bar d}0^{ud}1^{\bar u \bar u})
\nonumber\\
&&\nonumber\\
&+&2\, \alpha_2^{0^{ud}0^{\bar u \bar d}}
I_6(1^{uu}0^{\bar u \bar d}0^{ud}0^{\bar u \bar d})
+2\, \alpha_3^{1^{\bar u \bar u}1^{u \bar d}0^{ud}}
I_8(1^{uu}0^{\bar u \bar d}0^{ud}1^{u \bar d}1^{\bar u \bar u})
\nonumber\\
&&\nonumber\\
&+&2\, \alpha_3^{0^{ud}0^{u \bar u}0^{\bar u \bar d}}
I_8(1^{uu}0^{\bar u \bar d}0^{ud}0^{u \bar u}0^{\bar u \bar d})
\nonumber\\
&&\nonumber\\
\alpha_2^{1^{\bar u \bar u}0^{ud}}&=&\lambda
+\alpha_1^{1^{uu}} I_4(0^{ud}1^{\bar u \bar u}1^{uu})
+2\, \alpha_1^{1^{u \bar u}} I_3(1^{\bar u \bar u}0^{ud}1^{u \bar u})
+2\, \alpha_1^{1^{d \bar u}} I_3(1^{\bar u \bar u}0^{ud}1^{d \bar u})
+\alpha_1^{0^{ud}} I_4(0^{ud}1^{\bar u \bar u}0^{ud})
\nonumber\\
&&\nonumber\\
&+&2\, \alpha_1^{0^{\bar u \bar d}}
I_4(1^{\bar u \bar u}0^{ud}0^{\bar u \bar d})
+2\, \alpha_1^{0^{u \bar u}} I_3(1^{\bar u \bar u}0^{ud}0^{u \bar u})
+2\, \alpha_1^{0^{d \bar u}} I_3(1^{\bar u \bar u}0^{ud}0^{d \bar u})
\nonumber\\
&&\nonumber\\
&+&2\, \alpha_2^{1^{uu}0^{\bar u \bar d}}
I_6(1^{\bar u \bar u}0^{ud}0^{\bar u \bar d}1^{uu})
+2\, \alpha_2^{0^{ud}0^{\bar u \bar d}}
I_6(1^{\bar u \bar u}0^{ud}0^{\bar u \bar d}0^{ud})
+2\, \alpha_3^{1^{uu}1^{d \bar u}0^{\bar u \bar d}}
I_8(1^{\bar u \bar u}0^{ud}0^{\bar u \bar d}1^{d \bar u}1^{uu})
\nonumber\\
&&\nonumber\\
&+&2\, \alpha_3^{0^{ud}0^{u \bar u}0^{\bar u \bar d}}
I_8(1^{\bar u \bar u}0^{ud}0^{\bar u \bar d}0^{u \bar u}0^{ud})
\nonumber\\
&&\nonumber\\
\alpha_2^{0^{ud}0^{\bar u \bar d}}&=&\lambda
+\alpha_1^{1^{uu}} I_4(0^{ud}0^{\bar u \bar d}1^{uu})
+\alpha_1^{1^{\bar u \bar u}} I_4(0^{\bar u \bar d}0^{ud}1^{\bar u \bar u})
+\alpha_1^{1^{u \bar u}} I_3(0^{ud}0^{\bar u \bar d}1^{u \bar u})
+\alpha_1^{1^{u \bar d}} I_3(0^{ud}0^{\bar u \bar d}1^{u \bar d})
\nonumber\\
&&\nonumber\\
&+&\alpha_1^{1^{d \bar u}} I_3(0^{ud}0^{\bar u \bar d}1^{d \bar u})
+\alpha_1^{1^{d \bar d}} I_3(0^{ud}0^{\bar u \bar d}1^{d \bar d})
+\alpha_1^{0^{ud}} I_4(0^{ud}0^{\bar u \bar d}0^{ud})
+\alpha_1^{0^{\bar u \bar d}} I_4(0^{\bar u \bar d}0^{ud}0^{\bar u \bar d})
\nonumber\\
&&\nonumber\\
&+&\alpha_1^{0^{u \bar u}} I_3(0^{ud}0^{\bar u \bar d}0^{u \bar u})
+\alpha_1^{0^{u \bar d}} I_3(0^{ud}0^{\bar u \bar d}0^{u \bar d})
+\alpha_1^{0^{d \bar u}} I_3(0^{ud}0^{\bar u \bar d}0^{d \bar u})
+\alpha_1^{0^{d \bar d}} I_3(0^{ud}0^{\bar u \bar d}0^{d \bar d})
\nonumber\\
&&\nonumber\\
&+&\alpha_2^{1^{uu}1^{\bar u \bar u}}
I_6(0^{ud}0^{\bar u \bar d}1^{uu}1^{\bar u \bar u})
+\alpha_2^{1^{uu}0^{\bar u \bar d}}
I_6(0^{ud}0^{\bar u \bar d}1^{uu}0^{\bar u \bar d})
+\alpha_2^{1^{\bar u \bar u}0^{ud}}
I_6(0^{ud}0^{\bar u \bar d}0^{ud}1^{\bar u \bar u})
\nonumber\\
&&\nonumber\\
&+&\alpha_2^{0^{ud}0^{\bar u \bar d}}
I_6(0^{ud}0^{\bar u \bar d}0^{ud}0^{\bar u \bar d})
+\alpha_3^{1^{uu}1^{\bar u \bar u}0^{d \bar d}}
I_8(0^{ud}0^{\bar u \bar d}1^{uu}0^{d \bar d}1^{\bar u \bar u})
+\alpha_3^{1^{uu}0^{d \bar u}0^{\bar u \bar d}}
I_8(0^{ud}0^{\bar u \bar d}1^{uu}0^{d \bar u}0^{\bar u \bar d})
\nonumber\\
&&\nonumber\\
&+&\alpha_3^{1^{\bar u \bar u}1^{u \bar d}0^{ud}}
I_8(0^{ud}0^{\bar u \bar d}0^{ud}1^{u \bar d}1^{\bar u \bar u})
+\alpha_3^{0^{ud}0^{u \bar u}0^{\bar u \bar d}}
I_8(0^{ud}0^{\bar u \bar d}0^{ud}0^{u \bar u}0^{\bar u \bar d})
\nonumber\\
&&\nonumber\\
\alpha_3^{1^{uu}1^{\bar u \bar u}0^{d \bar d}}&=&\lambda
+2\, \alpha_1^{1^{u \bar d}}
I_9(1^{uu}0^{d \bar d}1^{\bar u \bar u}1^{u \bar d})
+2\, \alpha_1^{1^{d \bar u}}
I_9(1^{\bar u \bar u}0^{d \bar d}1^{uu}1^{d \bar u})
+2\, \alpha_1^{0^{ud}}
I_9(1^{uu}0^{d \bar d}1^{\bar u \bar u}0^{ud})
\nonumber\\
&&\nonumber\\
&+&2\, \alpha_1^{0^{\bar u \bar d}}
I_9(1^{\bar u \bar u}0^{d \bar d}1^{uu}0^{\bar u \bar d})
+4\, \alpha_1^{0^{u \bar u}}
I_9(1^{uu}1^{\bar u \bar u}0^{d \bar d}0^{u \bar u})
+2\, \alpha_1^{0^{u \bar d}}
I_9(1^{uu}0^{d \bar d}1^{\bar u \bar u}0^{u \bar d})
\nonumber\\
&&\nonumber\\
&+&2\, \alpha_1^{0^{d \bar u}}
I_9(1^{\bar u \bar u}0^{d \bar d}1^{uu}0^{d \bar u})
+4\, \alpha_2^{0^{ud}0^{\bar u \bar d}}
I_{10}(1^{uu}0^{d \bar d}1^{\bar u \bar u}0^{ud}0^{\bar u \bar d})
\nonumber\\
&&\nonumber\\
\alpha_3^{1^{uu}1^{d \bar u}0^{\bar u \bar d}}&=&\lambda
+\alpha_1^{1^{\bar u \bar u}}
I_9(1^{d \bar u}0^{\bar u \bar d}1^{uu}1^{\bar u \bar u})
+2\, \alpha_1^{1^{u \bar u}}
I_9(1^{uu}0^{\bar u \bar d}1^{d \bar u}1^{u \bar u})
+2\, \alpha_1^{1^{u \bar d}}
I_9(1^{uu}0^{\bar u \bar d}1^{d \bar u}1^{u \bar d})
\nonumber\\
&&\nonumber\\
&+&\alpha_1^{1^{d \bar u}}
I_9(1^{d \bar u}0^{\bar u \bar d}1^{uu}1^{d \bar u})
+\alpha_1^{1^{d \bar d}}
I_9(1^{d \bar u}0^{\bar u \bar d}1^{uu}1^{d \bar d})
+2\, \alpha_1^{0^{ud}}
I_9(1^{uu}1^{d \bar u}0^{\bar u \bar d}0^{ud})
\nonumber\\
&&\nonumber\\
&+&\alpha_1^{0^{\bar u \bar d}}
I_9(1^{d \bar u}0^{\bar u \bar d}1^{uu}0^{\bar u \bar d})
+\alpha_1^{0^{u \bar u}}
(2\, I_9(1^{uu}1^{d \bar u}0^{\bar u \bar d}0^{u \bar u})
+2\, I_9(1^{uu}0^{\bar u \bar d}1^{d \bar u}0^{u \bar u}))
\nonumber\\
&&\nonumber\\
&+&2\, \alpha_1^{0^{u \bar d}}
I_9(1^{uu}0^{\bar u \bar d}1^{d \bar u}0^{u \bar d})
+\alpha_1^{0^{d \bar u}}
I_9(1^{d \bar u}0^{\bar u \bar d}1^{uu}0^{d \bar u})
+\alpha_1^{0^{d \bar d}}
I_9(1^{d \bar u}0^{\bar u \bar d}1^{uu}0^{d \bar d})
\nonumber\\
&&\nonumber\\
&+&2\, \alpha_2^{1^{\bar u \bar u}0^{ud}}
I_{10}(1^{uu}1^{d \bar u}0^{\bar u \bar d}0^{ud}1^{\bar u \bar u})
+2\, \alpha_2^{0^{ud}0^{\bar u \bar d}}
I_{10}(1^{uu}1^{d \bar u}0^{\bar u \bar d}0^{ud}0^{\bar u \bar d})
\nonumber\\
&&\nonumber\\
\alpha_3^{1^{\bar u \bar u}1^{u \bar d}0^{ud}}&=&\lambda
+\alpha_1^{1^{uu}} I_9(1^{u \bar d}0^{ud}1^{\bar u \bar u}1^{uu})
+2\, \alpha_1^{1^{u \bar u}}
I_9(1^{\bar u \bar u}0^{ud}1^{u \bar d}1^{u \bar u})
+\alpha_1^{1^{u \bar d}}
I_9(1^{u \bar d}0^{ud}1^{\bar u \bar u}1^{u \bar d})
\nonumber\\
&&\nonumber\\
&+&2\, \alpha_1^{1^{d \bar u}}
I_9(1^{\bar u \bar u}0^{ud}1^{u \bar d}1^{d \bar u})
+\alpha_1^{1^{d \bar d}}
I_9(1^{u \bar d}0^{ud}1^{\bar u \bar u}1^{d \bar d})
+\alpha_1^{0^{ud}}
I_9(1^{u \bar d}0^{ud}1^{\bar u \bar u}0^{ud})
\nonumber\\
&&\nonumber\\
&+&2\, \alpha_1^{0^{\bar u \bar d}}
I_9(1^{\bar u \bar u}1^{u \bar d}0^{ud}0^{\bar u \bar d})
+\alpha_1^{0^{u \bar u}}
(2\, I_9(1^{\bar u \bar u}1^{u \bar d}0^{ud}0^{u \bar u})
+2\, I_9(1^{\bar u \bar u}0^{ud}1^{u \bar d}0^{u \bar u}))
\nonumber\\
&&\nonumber\\
&+&\alpha_1^{0^{u \bar d}}
I_9(1^{u \bar d}0^{ud}1^{\bar u \bar u}0^{u \bar d})
+2\, \alpha_1^{0^{d \bar u}}
I_9(1^{\bar u \bar u}0^{ud}1^{u \bar d}0^{d \bar u})
+\alpha_1^{0^{d \bar d}}
I_9(1^{u \bar d}0^{ud}1^{\bar u \bar u}0^{d \bar d})
\nonumber\\
&&\nonumber\\
&+&2\, \alpha_2^{1^{uu}0^{\bar u \bar d}}
I_{10}(1^{\bar u \bar u}1^{u \bar d}0^{ud}0^{\bar u \bar d}1^{uu})
+2\, \alpha_2^{0^{ud}0^{\bar u \bar d}}
I_{10}(1^{\bar u \bar u}1^{u \bar d}0^{ud}0^{\bar u \bar d}0^{ud})
\nonumber\\
&&\nonumber\\
\alpha_3^{0^{ud}0^{u \bar u}0^{\bar u \bar d}}&=&\lambda
+\alpha_1^{1^{uu}} I_9(0^{ud}0^{u \bar u}0^{\bar u \bar d}1^{uu})
+\alpha_1^{1^{\bar u \bar u}}
I_9(0^{\bar u \bar d}0^{u \bar u}0^{ud}1^{\bar u \bar u})
+\alpha_1^{1^{u \bar u}}
(I_9(0^{ud}0^{u \bar u}0^{\bar u \bar d}1^{u \bar u})
\nonumber\\
&&\nonumber\\
&+&I_9(0^{\bar u \bar d}0^{u \bar u}0^{ud}1^{u \bar u})
+I_9(0^{ud}0^{\bar u \bar d}0^{u \bar u}1^{u \bar u}))
+\alpha_1^{1^{u \bar d}}
(I_9(0^{\bar u \bar d}0^{u \bar u}0^{ud}1^{u \bar d})
+I_9(0^{ud}0^{\bar u \bar d}0^{u \bar u}1^{u \bar d}))
\nonumber\\
&&\nonumber\\
&+&\alpha_1^{1^{d \bar u}}
(I_9(0^{ud}0^{u \bar u}0^{\bar u \bar d}1^{d \bar u})
+I_9(0^{ud}0^{\bar u \bar d}0^{u \bar u}1^{d \bar u}))
+\alpha_1^{1^{d \bar d}}
I_9(0^{ud}0^{\bar u \bar d}0^{u \bar u}1^{d \bar d})
+\alpha_1^{0^{ud}}
I_9(0^{ud}0^{u \bar u}0^{\bar u \bar d}0^{ud})
\nonumber\\
&&\nonumber\\
&+&\alpha_1^{0^{\bar u \bar d}}
I_9(0^{\bar u \bar d}0^{u \bar u}0^{ud}0^{\bar u \bar d})
+\alpha_1^{0^{u \bar u}}
(I_9(0^{ud}0^{u \bar u}0^{\bar u \bar d}0^{u \bar u})
+I_9(0^{\bar u \bar d}0^{u \bar u}0^{ud}0^{u \bar u})
+I_9(0^{ud}0^{\bar u \bar d}0^{u \bar u}0^{u \bar u}))
\nonumber\\
&&\nonumber\\
&+&\alpha_1^{0^{u \bar d}}
(I_9(0^{\bar u \bar d}0^{u \bar u}0^{ud}0^{u \bar d})
+I_9(0^{ud}0^{\bar u \bar d}0^{u \bar u}0^{u \bar d}))
+\alpha_1^{0^{d \bar u}}
(I_9(0^{ud}0^{u \bar u}0^{\bar u \bar d}0^{d \bar u})
+I_9(0^{ud}0^{\bar u \bar d}0^{u \bar u}0^{d \bar u}))
\nonumber\\
&&\nonumber\\
&+&\alpha_1^{0^{d \bar d}}
I_9(0^{ud}0^{\bar u \bar d}0^{u \bar u}0^{d \bar d})
+\alpha_2^{1^{uu}1^{\bar u \bar u}}
I_{10}(0^{ud}0^{u \bar u}0^{\bar u \bar d}1^{uu}1^{\bar u \bar u})
+\alpha_2^{1^{uu}0^{\bar u \bar d}}
I_{10}(0^{ud}0^{u \bar u}0^{\bar u \bar d}1^{uu}0^{\bar u \bar d})
\nonumber\\
&&\nonumber\\
&+&\alpha_2^{1^{\bar u \bar u}0^{ud}}
I_{10}(0^{ud}0^{u \bar u}0^{\bar u \bar d}0^{ud}1^{\bar u \bar u})
+\alpha_2^{0^{ud}0^{\bar u \bar d}}
I_{10}(0^{ud}0^{u \bar u}0^{\bar u \bar d}0^{ud}0^{\bar u \bar d})
\nonumber
\end{eqnarray}

\newpage

\noindent
Table I. $S$-wave baryon masses $M(J^p)$ $(GeV)$.

\vskip2.5ex

\noindent
\begin{tabular}{|ccp{4cm}ccc|}
\hline
 & $M(\frac{1}{2}^+)$ &  & & $M(\frac{3}{2}^+)$ &
\\[5pt]
\hline
$N$ &
\begin{tabular}{c}
$0.940$ \\
$0.940$
\end{tabular}
& $(0.940)$ & $\Delta$ &
\begin{tabular}{c}
$1.232$ \\
$1.232$
\end{tabular}
& $(1.232)$
\\[5pt]
$\Lambda$ &
\begin{tabular}{c}
$1.022$ \\
$1.098$
\end{tabular}
& $(1.116)$ & $\Sigma^*$ &
\begin{tabular}{c}
$1.377$ \\
$1.377$
\end{tabular}
& $(1.385)$
\\[5pt]
$\Sigma$ &
\begin{tabular}{c}
$1.050$ \\
$1.193$
\end{tabular}
& $(1.193)$ & $\Xi^*$ &
\begin{tabular}{c}
$1.524$ \\
$1.524$
\end{tabular}
& $(1.530)$
\\[5pt]
$\Xi$ &
\begin{tabular}{c}
$1.162$ \\
$1.325$
\end{tabular}
& $(1.315)$ & $\Omega$ &
\begin{tabular}{c}
$1.672$ \\
$1.672$
\end{tabular}
& $(1.672)$
\\[5pt]
\hline
\end{tabular}

\vskip12ex

Table II. S-wave baryonia masses. Parameters of model: cutoff
$\Lambda=11.0$, gluon coupling constant $g=0.314$.
Quark masses $m_{u,d}=410\, MeV$.

\vskip2ex

\begin{tabular}{|c|c|c|c|}
\hline
$I$ & Quark content (baryonia) & $J$ & Mass (MeV) \\[5pt]
\hline
$0$ &
\begin{tabular}{l}
$uuu \bar u \bar u \bar u$ ($\Delta \bar \Delta$),\\
$ddd \bar d \bar d \bar d$ ($\Delta \bar \Delta$)
\end{tabular}
&
\begin{tabular}{c}
$0$ \\
$1$ \\
$2$ \\
$3$
\end{tabular}
&
\begin{tabular}{c}
$1973$ \\
$1824$ \\
$1938$ \\
$2290$
\end{tabular}
\\
\cline{2-4}
&
\begin{tabular}{l}
$udd \bar u \bar d \bar d$
($\Delta \bar \Delta+\Delta \bar n+n \bar \Delta+n\bar n$),\\
 $uud \bar u \bar u \bar d$
($\Delta \bar \Delta+\Delta \bar p+p \bar \Delta+p\bar p$)
\end{tabular}
&
\begin{tabular}{c}
$0$ \\
$1$ \\
$2$ \\
$3$
\end{tabular}
&
\begin{tabular}{c}
$1835$ \\
$1784$ \\
$1851$ \\
$2455$
\end{tabular}
\\
\hline
$1$ &
\begin{tabular}{l}
$udd \bar d \bar d \bar d$ ($\Delta \bar \Delta+n \bar \Delta$),\\
$uuu \bar u \bar u \bar d$ ($\Delta \bar \Delta+\Delta \bar p$)
\end{tabular}
&
\begin{tabular}{c}
$0$ \\
$1$ \\
$2$ \\
$3$
\end{tabular}
&
\begin{tabular}{c}
$1928$ \\
$1770$ \\
$1857$ \\
$2395$
\end{tabular}
\\
\cline{2-4}
&
\begin{tabular}{l}
$uud \bar u \bar d \bar d$
($\Delta \bar \Delta+\Delta \bar n+p \bar \Delta+p\bar n$)
\end{tabular}
&
\begin{tabular}{c}
$0$ \\
$1$ \\
$2$ \\
$3$
\end{tabular}
&
\begin{tabular}{c}
$1835$ \\
$1784$ \\
$1851$ \\
$2455$
\end{tabular}
\\
\hline
$2$ &
\begin{tabular}{l}
$uud \bar d \bar d \bar d$ ($\Delta \bar \Delta+p \bar \Delta$),\\
$uuu \bar u \bar d \bar d$ ($\Delta \bar \Delta+\Delta \bar n$)
\end{tabular}
&
\begin{tabular}{c}
$0$ \\
$1$ \\
$2$ \\
$3$
\end{tabular}
&
\begin{tabular}{c}
$1928$ \\
$1770$ \\
$1857$ \\
$2395$
\end{tabular}
\\
\hline
$3$ &
\begin{tabular}{l}
$uuu \bar d \bar d \bar d$ ($\Delta \bar \Delta$)
\end{tabular}
&
\begin{tabular}{c}
$0$ \\
$1$ \\
$2$ \\
$3$
\end{tabular}
&
\begin{tabular}{c}
$2067$ \\
$1783$ \\
$1938$ \\
$2290$
\end{tabular}
\\
\hline
\end{tabular}

\newpage

Table III. $IJ=00$, $uud \bar u \bar u \bar d$ $(1835 \, MeV)$,
$\Lambda=11.0$, $g=0.314$.
\vskip2ex

\begin{tabular}{|c|c|}
\hline
 Subamplitudes & Contributions, percent \\
\hline
& \\
$A_1^{1^{uu}}$ & $3.7$ \\
$A_1^{1^{\bar u \bar u}}$ & $3.7$ \\
$A_1^{1^{u \bar u}}$ & $9.0$ \\
$A_1^{1^{u \bar d}}$ & $7.8$ \\
$A_1^{1^{d \bar u}}$ & $7.8$ \\
$A_1^{1^{d \bar d}}$ & $6.7$ \\
$A_1^{0^{ud}}$ & $3.6$ \\
$A_1^{0^{\bar u \bar d}}$ & $3.6$ \\
$A_1^{0^{u \bar u}}$ & $7.0$ \\
$A_1^{0^{u \bar d}}$ & $6.8$ \\
$A_1^{0^{d \bar u}}$ & $6.8$ \\
$A_1^{0^{d \bar d}}$ & $6.6$ \\
$A_2^{1^{uu}1^{\bar u \bar u}}$ & $2.4$ \\
$A_2^{1^{uu}0^{\bar u \bar d}}$ & $2.0$ \\
$A_2^{1^{\bar u \bar u}0^{ud}}$ & $2.6$ \\
$A_2^{0^{ud}0^{\bar u \bar d}}$ & $2.8$ \\
$A_3^{1^{uu}0^{d \bar d}1^{\bar u \bar u}}$ & $4.0$ \\
$A_3^{1^{uu}1^{d \bar u}0^{\bar u \bar d}}$ & $4.5$ \\
$A_3^{1^{\bar u \bar u}1^{u \bar d}0^{ud}}$ & $4.5$ \\
$A_3^{0^{ud}0^{u \bar u}0^{\bar u \bar d}}$ & $4.3$ \\
\hline
$\sum A_1$ & $73.0$ \\
$\sum A_2$ & $9.8$ \\
$\sum A_3$ & $17.2$ \\
\hline
\end{tabular}

\vskip16ex

Table IV. $IJ=33$, $uuu \bar d \bar d \bar d$ $(2290 \, MeV)$,
$\Lambda=11.0$, $g=0.314$.

\vskip2ex

\begin{tabular}{|c|c|}
\hline
 Subamplitudes & Contributions, percent \\
\hline
& \\
$A_1^{1^{uu}}$ & $9.9$ \\
$A_1^{1^{\bar d \bar d}}$ & $9.9$ \\
$A_1^{1^{u \bar d}}$ & $25.4$ \\
$A_2^{1^{uu}1^{\bar d \bar d}}$ & $14.5$ \\
$A_3^{1^{uu}1^{\bar d \bar d}1^{u \bar d}}$ & $40.3$ \\
\hline
$\sum A_1$ & $45.2$ \\
$\sum A_2$ & $14.5$ \\
$\sum A_3$ & $40.3$ \\
\hline
\end{tabular}

\newpage

\noindent
Table V. Vertex functions and Chew-Mandelstam coefficients.

\vskip3ex

\begin{tabular}{|c|c|c|c|c|}
\hline
$i$ & $G_i^2(s_{kl})$ & $\alpha_i$ & $\beta_i$ & $\delta_i$ \\
\hline
& & & & \\
$0^+$ diquark & $\frac{4g}{3}-\frac{8gm_{kl}^2}{(3s_{kl})}$
& $\frac{1}{2}$ & $-\frac{1}{2}\frac{(m_k-m_l)^2}{(m_k+m_l)^2}$ & $0$ \\
& & & & \\
$1^+$ diquark & $\frac{2g}{3}$ & $\frac{1}{3}$
& $\frac{4m_k m_l}{3(m_k+m_l)^2}-\frac{1}{6}$
& $-\frac{1}{6}\frac{(m_k-m_l)^2}{(m_k+m_l)^2}$ \\
& & & & \\
$0^-$ meson & $\frac{8g}{3}-\frac{16gm_{kl}^2}{(3s_{kl})}$
& $\frac{1}{2}$ & $-\frac{1}{2}\frac{(m_k-m_l)^2}{(m_k+m_l)^2}$ & $0$ \\
& & & & \\
$1^-$ meson & $\frac{4g}{3}$ & $\frac{1}{3}$
& $\frac{4m_k m_l}{3(m_k+m_l)^2}-\frac{1}{6}$
& $-\frac{1}{6}\frac{(m_k-m_l)^2}{(m_k+m_l)^2}$ \\
& & & & \\
\hline
\end{tabular}

\newpage

\begin{picture}(600,100)
\put(0,53){\line(1,0){20}}
\put(0,50){\line(1,0){20}}
\put(0,47){\line(1,0){20}}
\put(20,53){\vector(3,2){23}}
\put(20,50){\vector(1,0){25}}
\put(20,47){\vector(3,-2){23}}
\put(58,47){+}
\put(78,53){\line(1,0){19}}
\put(78,50){\line(1,0){21}}
\put(78,47){\line(1,0){20}}
\put(104,57){\circle{16}}
\put(98,47){\vector(3,-2){23}}
\put(111,61){\vector(1,2){11}}
\put(111,61){\vector(4,1){25}}
\put(145,47){+}
\put(165,53){\line(1,0){19}}
\put(165,50){\line(1,0){21}}
\put(165,47){\line(1,0){20}}
\put(191,57){\circle{16}}
\put(185,47){\vector(3,-2){23}}
\put(202.5,68.5){\circle{16}}
\put(214,80){\circle{16}}
\put(220,86){\vector(1,2){11}}
\put(220,86){\vector(4,1){25}}
\put(255,47){+}
\put(275,53){\line(1,0){19}}
\put(275,50){\line(1,0){21}}
\put(275,47){\line(1,0){20}}
\put(301,57){\circle{16}}
\put(295,47){\vector(4,-1){95}}
\put(312.5,68.5){\circle{16}}
\put(324,80){\circle{16}}
\put(330,86){\vector(1,2){11}}
\put(330,86){\vector(2,-3){50}}
\put(405,47){+}
\put(430,47){$\cdots$}
\put(15,15){a}
\put(95,15){b}
\put(185,15){c}
\put(330,15){d}
\end{picture}

\vskip30pt

\begin{picture}(600,100)
\put(0,53){\line(1,0){19}}
\put(0,50){\line(1,0){21}}
\put(0,47){\line(1,0){20}}
\put(26,57){\circle{16}}
\put(20,47){\vector(4,-1){73}}
\put(37.5,68.5){\circle{16}}
\put(49,80){\circle{16}}
\put(55,86){\vector(1,2){11}}
\put(55,86){\vector(2,-3){38}}
\put(101,27){\circle{16}}
\put(116,22){\circle{16}}
\put(131,17){\circle{16}}
\put(139,15){\vector(4,3){17}}
\put(139,15){\vector(1,-1){15}}
\put(177,47){+}
\put(200,53){\line(1,0){19}}
\put(200,50){\line(1,0){21}}
\put(200,47){\line(1,0){20}}
\put(226,57){\circle{16}}
\put(220,47){\vector(4,-1){73}}
\put(237.5,68.5){\circle{16}}
\put(249,80){\circle{16}}
\put(255,86){\vector(4,-1){96}}
\put(255,86){\vector(2,-3){38}}
\put(301,27){\circle{16}}
\put(316,22){\circle{16}}
\put(331,17){\circle{16}}
\put(339,15){\vector(1,4){11.7}}
\put(339,15){\vector(1,-1){15}}
\put(359,63){\circle{16}}
\put(374,70){\circle{16}}
\put(389,77){\circle{16}}
\put(397,80){\vector(1,1){15}}
\put(397,80){\vector(4,-3){16}}
\put(65,5){e}
\put(270,5){f}
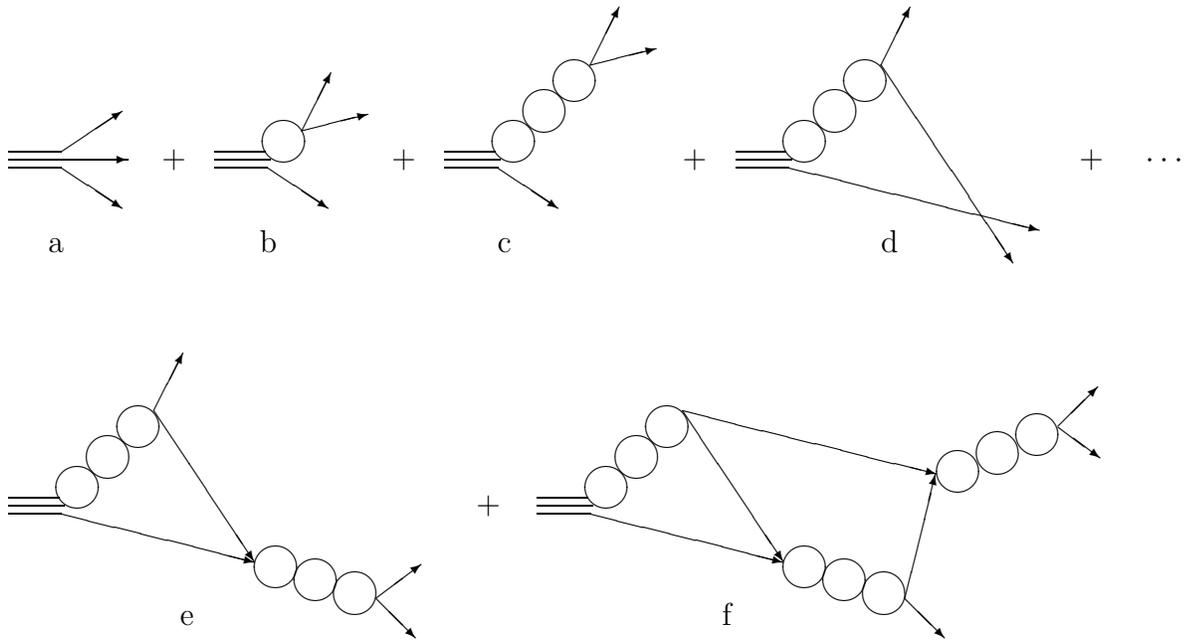
\put(0,-25){Fig. 1. Diagrams which correspond to a) production of three
quarks, b -- f) subsequent pair}
\put(0,-40){interaction.}
\end{picture}

\vskip150pt

\begin{picture}(600,100)
\put(0,53){\line(1,0){18}}
\put(0,50){\line(1,0){18}}
\put(0,47){\line(1,0){18}}
\put(30,50){\circle{25}}
\put(19,46){\line(1,1){15}}
\put(22,41){\line(1,1){17}}
\put(27.5,38.5){\line(1,1){14}}
\put(47,61){\circle{16}}
\put(54,65){\vector(1,1){18}}
\put(54,65){\vector(3,-1){22}}
\put(40,42){\vector(3,-2){24}}
\put(78,80){1}
\put(80,50){2}
\put(71,22){3}
\put(96,47){$=$}
\put(116,53){\line(1,0){19}}
\put(116,50){\line(1,0){21}}
\put(116,47){\line(1,0){20}}
\put(142,57){\circle{16}}
\put(136,47){\vector(3,-2){23}}
\put(149,61){\vector(1,2){11}}
\put(149,61){\vector(4,1){25}}
\put(167,80){1}
\put(181,62){2}
\put(167,24){3}
\put(193,47){+}
\put(215,53){\line(1,0){18}}
\put(215,50){\line(1,0){18}}
\put(215,47){\line(1,0){18}}
\put(245,50){\circle{25}}
\put(234,46){\line(1,1){15}}
\put(237,41){\line(1,1){17}}
\put(242.5,38.5){\line(1,1){14}}
\put(262,61){\circle{16}}
\put(269,65){\vector(1,1){18}}
\put(269,65){\vector(1,-1){23}}
\put(254,42){\vector(1,0){38}}
\put(300,40){\circle{16}}
\put(307,36){\vector(4,-1){25}}
\put(307,36){\vector(2,-3){14}}
\put(293,80){3}
\put(285,55){2}
\put(275,28){1}
\put(330,35){2}
\put(325,10){1}
\put(345,47){$+$}
\put(365,53){\line(1,0){18}}
\put(365,50){\line(1,0){18}}
\put(365,47){\line(1,0){18}}
\put(395,50){\circle{25}}
\put(384,46){\line(1,1){15}}
\put(387,41){\line(1,1){17}}
\put(392.5,38.5){\line(1,1){14}}
\put(412,61){\circle{16}}
\put(419,65){\vector(1,1){18}}
\put(419,65){\vector(1,-1){23}}
\put(404,42){\vector(1,0){38}}
\put(450,40){\circle{16}}
\put(457,36){\vector(4,-1){25}}
\put(457,36){\vector(2,-3){14}}
\put(443,80){3}
\put(435,55){1}
\put(425,28){2}
\put(480,35){1}
\put(475,10){2}
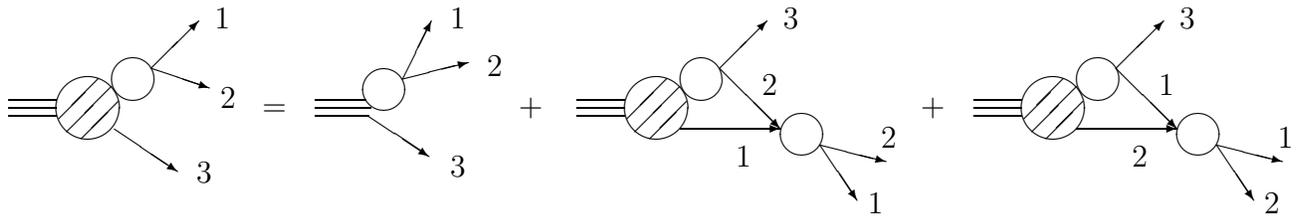
\put(0,-15){Fig. 2. Graphic representation of the equation for the amplitude
$A_1 (s,s_{12})$.}
\end{picture}

\newpage


\vskip30pt
\begin{picture}(600,100)
\put(0,45){\line(1,0){18}}
\put(0,47){\line(1,0){17.5}}
\put(0,49){\line(1,0){17}}
\put(0,51){\line(1,0){17}}
\put(0,53){\line(1,0){17.5}}
\put(0,55){\line(1,0){18}}
\put(30,50){\circle{25}}
\put(19,46){\line(1,1){15}}
\put(22,41){\line(1,1){17}}
\put(27.5,38.5){\line(1,1){14}}
\put(31,63){\vector(1,1){20}}
\put(37,60){\vector(1,1){20}}
\put(31,38){\vector(1,-1){20}}
\put(37,40){\vector(1,-1){20}}
\put(50,50){\circle{16}}
\put(58,50){\vector(3,2){22}}
\put(58,50){\vector(3,-2){22}}
\put(75,70){$u$}
\put(83,64){\small 1}
\put(75,23){$u$}
\put(83,32){\small 2}
\put(47,86){$\bar d$}
\put(38,82){\small 4}
\put(60,81){$u$}
\put(62,70){\small 3}
\put(39,10){$\bar d$}
\put(35,20){\small 6}
\put(60,13){$\bar d$}
\put(58,27){\small 5}
\put(43,47){\footnotesize $1^{uu}$}
\put(90,48){$=$}
\put(110,45){\line(1,0){18}}
\put(110,47){\line(1,0){17.5}}
\put(110,49){\line(1,0){17}}
\put(110,51){\line(1,0){17}}
\put(110,53){\line(1,0){17.5}}
\put(110,55){\line(1,0){18}}
\put(135,50){\circle{16}}
\put(128,55){\vector(3,2){22}}
\put(128,55){\vector(2,3){15}}
\put(128,45){\vector(3,-2){22}}
\put(128,45){\vector(2,-3){15}}
\put(143,50){\vector(3,1){22}}
\put(143,50){\vector(3,-1){22}}
\put(167,60){$u$}
\put(160,60){\small 1}
\put(167,33){$u$}
\put(160,34){\small 2}
\put(152,64){$u$}
\put(152,72){\small 3}
\put(144,81){$\bar d$}
\put(136,80){\small 4}
\put(152,21){$\bar d$}
\put(151,33){\small 5}
\put(142,12){$\bar d$}
\put(134,16){\small 6}
\put(128,47){\footnotesize $1^{uu}$}
\put(175,48){$+$}
\put(190,48){2}
\put(203,45){\line(1,0){18}}
\put(203,47){\line(1,0){17.5}}
\put(203,49){\line(1,0){17}}
\put(203,51){\line(1,0){17}}
\put(203,53){\line(1,0){17.5}}
\put(203,55){\line(1,0){18}}
\put(233,50){\circle{25}}
\put(222,46){\line(1,1){15}}
\put(225,41){\line(1,1){17}}
\put(230.5,38.5){\line(1,1){14}}
\put(249,62){\circle{16}}
\put(254.5,68.5){\vector(1,1){15}}
\put(254.5,68.5){\vector(1,-1){18}}
\put(245,50){\vector(1,0){28}}
\put(281,50){\circle{16}}
\put(289,50){\vector(3,1){22}}
\put(289,50){\vector(3,-1){22}}
\put(243,43){\vector(1,-1){15}}
\put(240.5,40.5){\vector(2,-3){12}}
\put(237.5,38){\vector(1,-3){6.5}}
\put(265,61){$u$}
\put(260,65){\small 1}
\put(260,40){$u$}
\put(253,40){\small 2}
\put(310,60){$u$}
\put(303,60){\small 1}
\put(310,33){$u$}
\put(302,33){\small 2}
\put(273,82){$u$}
\put(260,82){\small 3}
\put(264,21){$\bar d$}
\put(261,30){\small 4}
\put(253,10){$\bar d$}
\put(254,22){\small 5}
\put(242,7){$\bar d$}
\put(236,10){\small 6}
\put(274,47){\footnotesize $1^{uu}$}
\put(242,59){\footnotesize $1^{uu}$}
\put(320,48){$+$}
\put(335,48){6}
\put(348,45){\line(1,0){18}}
\put(348,47){\line(1,0){17.5}}
\put(348,49){\line(1,0){17}}
\put(348,51){\line(1,0){17}}
\put(348,53){\line(1,0){17.5}}
\put(348,55){\line(1,0){18}}
\put(378,50){\circle{25}}
\put(367,46){\line(1,1){15}}
\put(370,41){\line(1,1){17}}
\put(375.5,38.5){\line(1,1){14}}
\put(394,62){\circle{16}}
\put(399.5,68.5){\vector(1,1){15}}
\put(399.5,68.5){\vector(1,-1){18}}
\put(390,50){\vector(1,0){28}}
\put(426,50){\circle{16}}
\put(434,50){\vector(3,1){22}}
\put(434,50){\vector(3,-1){22}}
\put(388,43){\vector(1,-1){15}}
\put(385.5,40.5){\vector(2,-3){12}}
\put(382.5,38){\vector(1,-3){6.5}}
\put(410,61){$u$}
\put(405,65){\small 1}
\put(405,40){$u$}
\put(398,40){\small 2}
\put(455,60){$u$}
\put(448,60){\small 1}
\put(455,33){$u$}
\put(447,33){\small 2}
\put(418,82){$\bar d$}
\put(405,82){\small 3}
\put(406,21){$u$}
\put(406,30){\small 4}
\put(398,10){$\bar d$}
\put(399,22){\small 5}
\put(387,7){$\bar d$}
\put(381,10){\small 6}
\put(419,47){\footnotesize $1^{uu}$}
\put(387,59){\footnotesize $1^{u \bar d}$}
\end{picture}


\vskip30pt
\begin{picture}(600,100)
\put(0,45){\line(1,0){18}}
\put(0,47){\line(1,0){17.5}}
\put(0,49){\line(1,0){17}}
\put(0,51){\line(1,0){17}}
\put(0,53){\line(1,0){17.5}}
\put(0,55){\line(1,0){18}}
\put(30,50){\circle{25}}
\put(19,46){\line(1,1){15}}
\put(22,41){\line(1,1){17}}
\put(27.5,38.5){\line(1,1){14}}
\put(31,63){\vector(1,1){20}}
\put(37,60){\vector(1,1){20}}
\put(31,38){\vector(1,-1){20}}
\put(37,40){\vector(1,-1){20}}
\put(50,50){\circle{16}}
\put(58,50){\vector(3,2){22}}
\put(58,50){\vector(3,-2){22}}
\put(75,70){$\bar d$}
\put(83,64){\small 1}
\put(75,22){$\bar d$}
\put(83,32){\small 2}
\put(47,86){$u$}
\put(38,82){\small 4}
\put(60,81){$\bar d$}
\put(62,70){\small 3}
\put(39,10){$u$}
\put(35,20){\small 6}
\put(60,13){$u$}
\put(58,27){\small 5}
\put(43,47){\footnotesize $1^{\bar d \bar d}$}
\put(90,48){$=$}
\put(110,45){\line(1,0){18}}
\put(110,47){\line(1,0){17.5}}
\put(110,49){\line(1,0){17}}
\put(110,51){\line(1,0){17}}
\put(110,53){\line(1,0){17.5}}
\put(110,55){\line(1,0){18}}
\put(135,50){\circle{16}}
\put(128,55){\vector(3,2){22}}
\put(128,55){\vector(2,3){15}}
\put(128,45){\vector(3,-2){22}}
\put(128,45){\vector(2,-3){15}}
\put(143,50){\vector(3,1){22}}
\put(143,50){\vector(3,-1){22}}
\put(167,60){$\bar d$}
\put(160,60){\small 1}
\put(167,33){$\bar d$}
\put(160,34){\small 2}
\put(152,62){$\bar d$}
\put(152,74){\small 3}
\put(144,81){$u$}
\put(136,80){\small 4}
\put(152,22){$u$}
\put(151,33){\small 5}
\put(142,12){$u$}
\put(134,16){\small 6}
\put(128,47){\footnotesize $1^{\bar d \bar d}$}
\put(175,48){$+$}
\put(190,48){2}
\put(203,45){\line(1,0){18}}
\put(203,47){\line(1,0){17.5}}
\put(203,49){\line(1,0){17}}
\put(203,51){\line(1,0){17}}
\put(203,53){\line(1,0){17.5}}
\put(203,55){\line(1,0){18}}
\put(233,50){\circle{25}}
\put(222,46){\line(1,1){15}}
\put(225,41){\line(1,1){17}}
\put(230.5,38.5){\line(1,1){14}}
\put(249,62){\circle{16}}
\put(254.5,68.5){\vector(1,1){15}}
\put(254.5,68.5){\vector(1,-1){18}}
\put(245,50){\vector(1,0){28}}
\put(281,50){\circle{16}}
\put(289,50){\vector(3,1){22}}
\put(289,50){\vector(3,-1){22}}
\put(243,43){\vector(1,-1){15}}
\put(240.5,40.5){\vector(2,-3){12}}
\put(237.5,38){\vector(1,-3){6.5}}
\put(265,61){$\bar d$}
\put(260,65){\small 1}
\put(265,37){$\bar d$}
\put(256,40){\small 2}
\put(310,60){$\bar d$}
\put(303,60){\small 1}
\put(310,33){$\bar d$}
\put(302,33){\small 2}
\put(273,82){$\bar d$}
\put(260,82){\small 3}
\put(264,21){$u$}
\put(261,30){\small 4}
\put(253,10){$u$}
\put(254,22){\small 5}
\put(242,7){$u$}
\put(236,10){\small 6}
\put(274,47){\footnotesize $1^{\bar d \bar d}$}
\put(242,59){\footnotesize $1^{\bar d \bar d}$}
\put(320,48){$+$}
\put(335,48){6}
\put(348,45){\line(1,0){18}}
\put(348,47){\line(1,0){17.5}}
\put(348,49){\line(1,0){17}}
\put(348,51){\line(1,0){17}}
\put(348,53){\line(1,0){17.5}}
\put(348,55){\line(1,0){18}}
\put(378,50){\circle{25}}
\put(367,46){\line(1,1){15}}
\put(370,41){\line(1,1){17}}
\put(375.5,38.5){\line(1,1){14}}
\put(394,62){\circle{16}}
\put(399.5,68.5){\vector(1,1){15}}
\put(399.5,68.5){\vector(1,-1){18}}
\put(390,50){\vector(1,0){28}}
\put(426,50){\circle{16}}
\put(434,50){\vector(3,1){22}}
\put(434,50){\vector(3,-1){22}}
\put(388,43){\vector(1,-1){15}}
\put(385.5,40.5){\vector(2,-3){12}}
\put(382.5,38){\vector(1,-3){6.5}}
\put(410,61){$\bar d$}
\put(405,65){\small 1}
\put(410,37){$\bar d$}
\put(401,40){\small 2}
\put(455,60){$\bar d$}
\put(448,60){\small 1}
\put(455,33){$\bar d$}
\put(447,33){\small 2}
\put(418,82){$u$}
\put(405,82){\small 3}
\put(406,18){$\bar d$}
\put(406,30){\small 4}
\put(398,10){$u$}
\put(399,22){\small 5}
\put(387,7){$u$}
\put(381,10){\small 6}
\put(419,47){\footnotesize $1^{\bar d \bar d}$}
\put(387,59){\footnotesize $1^{u \bar d}$}
\end{picture}


\vskip30pt
\begin{picture}(600,100)
\put(0,45){\line(1,0){18}}
\put(0,47){\line(1,0){17.5}}
\put(0,49){\line(1,0){17}}
\put(0,51){\line(1,0){17}}
\put(0,53){\line(1,0){17.5}}
\put(0,55){\line(1,0){18}}
\put(30,50){\circle{25}}
\put(19,46){\line(1,1){15}}
\put(22,41){\line(1,1){17}}
\put(27.5,38.5){\line(1,1){14}}
\put(31,63){\vector(1,1){20}}
\put(37,60){\vector(1,1){20}}
\put(31,38){\vector(1,-1){20}}
\put(37,40){\vector(1,-1){20}}
\put(50,50){\circle{16}}
\put(58,50){\vector(3,2){22}}
\put(58,50){\vector(3,-2){22}}
\put(75,70){$u$}
\put(83,64){\small 1}
\put(75,23){$\bar d$}
\put(83,32){\small 2}
\put(47,86){$u$}
\put(38,82){\small 4}
\put(60,81){$u$}
\put(62,70){\small 3}
\put(39,10){$\bar d$}
\put(35,20){\small 6}
\put(60,13){$\bar d$}
\put(58,27){\small 5}
\put(43,47){\footnotesize $1^{u \bar d}$}
\put(90,48){$=$}
\put(110,45){\line(1,0){18}}
\put(110,47){\line(1,0){17.5}}
\put(110,49){\line(1,0){17}}
\put(110,51){\line(1,0){17}}
\put(110,53){\line(1,0){17.5}}
\put(110,55){\line(1,0){18}}
\put(135,50){\circle{16}}
\put(128,55){\vector(3,2){22}}
\put(128,55){\vector(2,3){15}}
\put(128,45){\vector(3,-2){22}}
\put(128,45){\vector(2,-3){15}}
\put(143,50){\vector(3,1){22}}
\put(143,50){\vector(3,-1){22}}
\put(167,60){$u$}
\put(160,60){\small 1}
\put(167,33){$\bar d$}
\put(160,34){\small 2}
\put(152,64){$u$}
\put(152,72){\small 3}
\put(144,81){$u$}
\put(136,80){\small 4}
\put(152,21){$\bar d$}
\put(151,33){\small 5}
\put(142,12){$\bar d$}
\put(134,16){\small 6}
\put(128,47){\footnotesize $1^{u \bar d}$}
\put(175,48){$+$}
\put(190,48){2}
\put(203,45){\line(1,0){18}}
\put(203,47){\line(1,0){17.5}}
\put(203,49){\line(1,0){17}}
\put(203,51){\line(1,0){17}}
\put(203,53){\line(1,0){17.5}}
\put(203,55){\line(1,0){18}}
\put(233,50){\circle{25}}
\put(222,46){\line(1,1){15}}
\put(225,41){\line(1,1){17}}
\put(230.5,38.5){\line(1,1){14}}
\put(249,62){\circle{16}}
\put(254.5,68.5){\vector(1,1){15}}
\put(254.5,68.5){\vector(1,-1){18}}
\put(245,50){\vector(1,0){28}}
\put(281,50){\circle{16}}
\put(289,50){\vector(3,1){22}}
\put(289,50){\vector(3,-1){22}}
\put(243,43){\vector(1,-1){15}}
\put(240.5,40.5){\vector(2,-3){12}}
\put(237.5,38){\vector(1,-3){6.5}}
\put(265,61){$u$}
\put(260,65){\small 1}
\put(260,38){$\bar d$}
\put(253,40){\small 2}
\put(310,60){$u$}
\put(303,60){\small 1}
\put(310,33){$\bar d$}
\put(302,33){\small 2}
\put(273,82){$u$}
\put(260,82){\small 3}
\put(261,21){$u$}
\put(261,30){\small 4}
\put(253,10){$\bar d$}
\put(254,22){\small 5}
\put(242,7){$\bar d$}
\put(236,10){\small 6}
\put(274,47){\footnotesize $1^{u \bar d}$}
\put(242,59){\footnotesize $1^{uu}$}
\put(320,48){$+$}
\put(335,48){2}
\put(348,45){\line(1,0){18}}
\put(348,47){\line(1,0){17.5}}
\put(348,49){\line(1,0){17}}
\put(348,51){\line(1,0){17}}
\put(348,53){\line(1,0){17.5}}
\put(348,55){\line(1,0){18}}
\put(378,50){\circle{25}}
\put(367,46){\line(1,1){15}}
\put(370,41){\line(1,1){17}}
\put(375.5,38.5){\line(1,1){14}}
\put(394,62){\circle{16}}
\put(399.5,68.5){\vector(1,1){15}}
\put(399.5,68.5){\vector(1,-1){18}}
\put(390,50){\vector(1,0){28}}
\put(426,50){\circle{16}}
\put(434,50){\vector(3,1){22}}
\put(434,50){\vector(3,-1){22}}
\put(388,43){\vector(1,-1){15}}
\put(385.5,40.5){\vector(2,-3){12}}
\put(382.5,38){\vector(1,-3){6.5}}
\put(410,61){$u$}
\put(405,65){\small 1}
\put(405,38){$\bar d$}
\put(398,40){\small 2}
\put(455,60){$u$}
\put(448,60){\small 1}
\put(455,33){$\bar d$}
\put(447,33){\small 2}
\put(418,82){$\bar d$}
\put(405,82){\small 3}
\put(406,21){$u$}
\put(406,30){\small 4}
\put(398,10){$u$}
\put(399,22){\small 5}
\put(387,7){$\bar d$}
\put(381,10){\small 6}
\put(419,47){\footnotesize $1^{u \bar d}$}
\put(387,59){\footnotesize $1^{u \bar d}$}
\end{picture}

\vskip30pt
\begin{picture}(600,100)
\put(90,48){$+$}
\put(105,48){2}
\put(118,45){\line(1,0){18}}
\put(118,47){\line(1,0){17.5}}
\put(118,49){\line(1,0){17}}
\put(118,51){\line(1,0){17}}
\put(118,53){\line(1,0){17.5}}
\put(118,55){\line(1,0){18}}
\put(148,50){\circle{25}}
\put(137,46){\line(1,1){15}}
\put(140,41){\line(1,1){17}}
\put(145.5,38.5){\line(1,1){14}}
\put(164,62){\circle{16}}
\put(169.5,68.5){\vector(1,1){15}}
\put(169.5,68.5){\vector(1,-1){18}}
\put(160,50){\vector(1,0){28}}
\put(196,50){\circle{16}}
\put(204,50){\vector(3,1){22}}
\put(204,50){\vector(3,-1){22}}
\put(158,43){\vector(1,-1){15}}
\put(155.5,40.5){\vector(2,-3){12}}
\put(152.5,38){\vector(1,-3){6.5}}
\put(180,61){$\bar d$}
\put(175,65){\small 1}
\put(175,40){$u$}
\put(168,40){\small 2}
\put(225,60){$u$}
\put(218,60){\small 1}
\put(225,33){$\bar d$}
\put(217,33){\small 2}
\put(188,82){$u$}
\put(175,82){\small 3}
\put(176,21){$u$}
\put(176,30){\small 4}
\put(168,10){$\bar d$}
\put(169,22){\small 5}
\put(157,7){$\bar d$}
\put(151,10){\small 6}
\put(189,47){\footnotesize $1^{u \bar d}$}
\put(157,59){\footnotesize $1^{u \bar d}$}
\put(235,48){$+$}
\put(250,48){2}
\put(263,45){\line(1,0){18}}
\put(263,47){\line(1,0){17.5}}
\put(263,49){\line(1,0){17}}
\put(263,51){\line(1,0){17}}
\put(263,53){\line(1,0){17.5}}
\put(263,55){\line(1,0){18}}
\put(293,50){\circle{25}}
\put(282,46){\line(1,1){15}}
\put(285,41){\line(1,1){17}}
\put(290.5,38.5){\line(1,1){14}}
\put(309,62){\circle{16}}
\put(314.5,68.5){\vector(1,1){15}}
\put(314.5,68.5){\vector(1,-1){18}}
\put(305,50){\vector(1,0){28}}
\put(341,50){\circle{16}}
\put(349,50){\vector(3,1){22}}
\put(349,50){\vector(3,-1){22}}
\put(303,43){\vector(1,-1){15}}
\put(300.5,40.5){\vector(2,-3){12}}
\put(297.5,38){\vector(1,-3){6.5}}
\put(325,61){$\bar d$}
\put(320,65){\small 1}
\put(320,40){$u$}
\put(313,40){\small 2}
\put(370,60){$u$}
\put(363,60){\small 1}
\put(370,33){$\bar d$}
\put(362,33){\small 2}
\put(333,82){$\bar d$}
\put(320,82){\small 3}
\put(321,21){$u$}
\put(321,30){\small 4}
\put(313,10){$u$}
\put(314,22){\small 5}
\put(302,7){$\bar d$}
\put(296,10){\small 6}
\put(334,47){\footnotesize $1^{u \bar d}$}
\put(302,59){\footnotesize $1^{\bar d \bar d}$}
\end{picture}

\vskip30pt
\begin{picture}(600,100)
\put(90,48){$+$}
\put(106,48){4}
\put(120,45){\line(1,0){18}}
\put(120,47){\line(1,0){17.5}}
\put(120,49){\line(1,0){17}}
\put(120,51){\line(1,0){17}}
\put(120,53){\line(1,0){17.5}}
\put(120,55){\line(1,0){18}}
\put(150,50){\circle{25}}
\put(139,46){\line(1,1){15}}
\put(142,41){\line(1,1){17}}
\put(147.5,38.5){\line(1,1){14}}
\put(151,63){\vector(1,1){20}}
\put(151,38){\vector(1,-1){20}}
\put(167.5,60){\circle{16}}
\put(167.5,40){\circle{16}}
\put(172,67){\vector(1,1){18}}
\put(172,33){\vector(1,-1){18}}
\put(172,67){\vector(1,-1){17}}
\put(172,33){\vector(1,1){17}}
\put(197,50){\circle{16}}
\put(205,50){\vector(1,1){18}}
\put(205,50){\vector(1,-1){18}}
\put(185,59){$u$}
\put(179,64){\small 1}
\put(181,31){$\bar d$}
\put(179,45){\small 2}
\put(226,65){$u$}
\put(212,65){\small 1}
\put(226,27){$\bar d$}
\put(212,27){\small 2}
\put(194,85){$u$}
\put(183,85){\small 3}
\put(194,10){$\bar d$}
\put(182,10){\small 4}
\put(173,83){$u$}
\put(160,83){\small 5}
\put(173,10){$\bar d$}
\put(160,12){\small 6}
\put(160.5,57){\footnotesize $1^{uu}$}
\put(160.5,37){\footnotesize $1^{\bar d \bar d}$}
\put(190,47){\footnotesize $1^{u \bar d}$}
\end{picture}


\vskip60pt
\begin{picture}(600,100)
\put(0,45){\line(1,0){18}}
\put(0,47){\line(1,0){17.5}}
\put(0,49){\line(1,0){17}}
\put(0,51){\line(1,0){17}}
\put(0,53){\line(1,0){17.5}}
\put(0,55){\line(1,0){18}}
\put(30,50){\circle{25}}
\put(19,46){\line(1,1){15}}
\put(22,41){\line(1,1){17}}
\put(27.5,38.5){\line(1,1){14}}
\put(31,63){\vector(1,1){20}}
\put(31,38){\vector(1,-1){20}}
\put(47.5,60){\circle{16}}
\put(47.5,40){\circle{16}}
\put(55,64){\vector(3,2){18}}
\put(55,36){\vector(3,-2){18}}
\put(55,64){\vector(3,-2){18}}
\put(55,36){\vector(3,2){18}}
\put(78,75){$u$}
\put(63,75){\small 1}
\put(78,53){$u$}
\put(70,58){\small 2}
\put(78,41){$\bar d$}
\put(70,36){\small 3}
\put(78,18){$\bar d$}
\put(63,18){\small 4}
\put(54,80){$u$}
\put(41,80){\small 5}
\put(54,13){$\bar d$}
\put(41,13){\small 6}
\put(40.5,56){\footnotesize $1^{uu}$}
\put(40.5,36){\footnotesize $1^{\bar d \bar d}$}
\put(90,48){$=$}
\put(110,45){\line(1,0){19}}
\put(110,47){\line(1,0){21}}
\put(110,49){\line(1,0){23}}
\put(110,51){\line(1,0){23}}
\put(110,53){\line(1,0){21}}
\put(110,55){\line(1,0){19}}
\put(140,60){\circle{16}}
\put(140,40){\circle{16}}
\put(147.5,64){\vector(3,2){18}}
\put(147.5,36){\vector(3,-2){18}}
\put(147.5,64){\vector(3,-2){18}}
\put(147.5,36){\vector(3,2){18}}
\put(128,55){\vector(1,3){11}}
\put(128,45){\vector(1,-3){11}}
\put(170,75){$u$}
\put(155,75){\small 1}
\put(170,53){$u$}
\put(159,59){\small 2}
\put(170,41){$\bar d$}
\put(159,35){\small 3}
\put(170,18){$\bar d$}
\put(155,18){\small 4}
\put(143,86){$u$}
\put(128,84){\small 5}
\put(143,08){$\bar d$}
\put(128,10){\small 6}
\put(133,57){\footnotesize $1^{uu}$}
\put(133,37){\footnotesize $1^{\bar d \bar d}$}
\put(183,48){$+$}
\put(199,48){4}
\put(212,45){\line(1,0){18}}
\put(212,47){\line(1,0){17.5}}
\put(212,49){\line(1,0){17}}
\put(212,51){\line(1,0){17}}
\put(212,53){\line(1,0){17.5}}
\put(212,55){\line(1,0){18}}
\put(242,50){\circle{25}}
\put(231,46){\line(1,1){15}}
\put(234,41){\line(1,1){17}}
\put(239.5,38.5){\line(1,1){14}}
\put(243,63){\vector(1,1){20}}
\put(243,38){\vector(1,-1){20}}
\put(266,80){$u$}
\put(252,80){\small 5}
\put(266,13){$\bar d$}
\put(252,13){\small 6}
\put(262,50){\circle{16}}
\put(270,50){\vector(3,2){17}}
\put(270,50){\vector(3,-2){17}}
\put(247,61){\vector(1,0){40}}
\put(247,39){\vector(1,0){40}}
\put(295,60){\circle{16}}
\put(295,40){\circle{16}}
\put(303,61){\vector(3,1){20}}
\put(303,61){\vector(3,-1){20}}
\put(303,39){\vector(3,1){20}}
\put(303,39){\vector(3,-1){20}}
\put(328,70){$u$}
\put(313,70){\small 1}
\put(328,53){$u$}
\put(310,51){\small 2}
\put(328,41){$\bar d$}
\put(310,44){\small 3}
\put(328,24){$\bar d$}
\put(313,24){\small 4}
\put(270,66){$u$}
\put(260,65){\small 1}
\put(275,49){\small $u$}
\put(270,54){\small 2}
\put(280,45){\small $\bar d$}
\put(270,40){\small 3}
\put(270,27){$\bar d$}
\put(260,28){\small 4}
\put(255,47){\footnotesize $1^{u \bar d}$}
\put(288,57){\footnotesize $1^{uu}$}
\put(288,37){\footnotesize $1^{\bar d \bar d}$}
\put(341,48){$+$}
\put(355,48){2}
\put(368,45){\line(1,0){18}}
\put(368,47){\line(1,0){17.5}}
\put(368,49){\line(1,0){17}}
\put(368,51){\line(1,0){17}}
\put(368,53){\line(1,0){17.5}}
\put(368,55){\line(1,0){18}}
\put(398,50){\circle{25}}
\put(387,46){\line(1,1){15}}
\put(390,41){\line(1,1){17}}
\put(395.5,38.5){\line(1,1){14}}
\put(414,62){\circle{16}}
\put(419.5,68.5){\vector(1,1){15}}
\put(419.5,68.5){\vector(1,-1){18}}
\put(410,50){\vector(1,0){28}}
\put(446,50){\circle{16}}
\put(454,50){\vector(3,1){22}}
\put(454,50){\vector(3,-1){22}}
\put(430,61){$u$}
\put(425,65){\small 1}
\put(427,41){$u$}
\put(425,52){\small 2}
\put(470,62){$u$}
\put(460,59){\small 1}
\put(470,30){$u$}
\put(460,34){\small 2}
\put(436,85){$u$}
\put(424,83){\small 5}
\put(414,37){\circle{16}}
\put(421,32){\vector(3,-1){20}}
\put(421,32){\vector(2,-3){12}}
\put(399,38){\vector(1,-3){8}}
\put(444,22){$\bar d$}
\put(434,32){\small 3}
\put(435,7){$\bar d$}
\put(425,7){\small 4}
\put(409,7){$\bar d$}
\put(397,9){\small 6}
\put(407,59){\footnotesize $1^{uu}$}
\put(439,47){\footnotesize $1^{uu}$}
\put(407,34){\footnotesize $1^{\bar d \bar d}$}
\end{picture}

\vskip60pt
\begin{picture}(600,60)
\put(90,48){$+$}
\put(107,48){2}
\put(125,45){\line(1,0){18}}
\put(125,47){\line(1,0){17.5}}
\put(125,49){\line(1,0){17}}
\put(125,51){\line(1,0){17}}
\put(125,53){\line(1,0){17.5}}
\put(125,55){\line(1,0){18}}
\put(155,50){\circle{25}}
\put(144,46){\line(1,1){15}}
\put(147,41){\line(1,1){17}}
\put(152.5,38.5){\line(1,1){14}}
\put(171,62){\circle{16}}
\put(176.5,68.5){\vector(1,1){15}}
\put(176.5,68.5){\vector(1,-1){18}}
\put(167,50){\vector(1,0){28}}
\put(203,50){\circle{16}}
\put(211,50){\vector(3,1){22}}
\put(211,50){\vector(3,-1){22}}
\put(187,61){$\bar d$}
\put(182,65){\small 1}
\put(184,38){$\bar d$}
\put(182,52){\small 2}
\put(227,62){$\bar d$}
\put(217,59){\small 1}
\put(227,30){$\bar d$}
\put(217,34){\small 2}
\put(193,85){$\bar d$}
\put(181,83){\small 5}
\put(171,37){\circle{16}}
\put(178,32){\vector(3,-1){20}}
\put(178,32){\vector(2,-3){12}}
\put(156,38){\vector(1,-3){8}}
\put(201,22){$u$}
\put(191,32){\small 3}
\put(192,7){$u$}
\put(182,7){\small 4}
\put(166,7){$u$}
\put(158,9){\small 6}
\put(164,59){\footnotesize $1^{\bar d \bar d}$}
\put(196,47){\footnotesize $1^{\bar d \bar d}$}
\put(164,34){\footnotesize $1^{uu}$}
\put(250,48){$+$}
\put(267,48){4}
\put(285,45){\line(1,0){18}}
\put(285,47){\line(1,0){17.5}}
\put(285,49){\line(1,0){17}}
\put(285,51){\line(1,0){17}}
\put(285,53){\line(1,0){17.5}}
\put(285,55){\line(1,0){18}}
\put(315,50){\circle{25}}
\put(304,46){\line(1,1){15}}
\put(307,41){\line(1,1){17}}
\put(312.5,38.5){\line(1,1){14}}
\put(324,68){\circle{16}}
\put(324,32){\circle{16}}
\put(327,53){\vector(3,1){28}}
\put(327,47){\vector(3,-1){28}}
\put(332,70){\vector(3,2){21}}
\put(332,70){\vector(3,-1){23}}
\put(332,30){\vector(3,1){23}}
\put(332,30){\vector(3,-2){21}}
\put(363,60){\circle{16}}
\put(363,40){\circle{16}}
\put(372,60){\vector(3,2){21}}
\put(372,60){\vector(3,-1){23}}
\put(372,40){\vector(3,1){23}}
\put(372,40){\vector(3,-2){21}}
\put(399,71){$u$}
\put(384,74){\small 1}
\put(400,53){$u$}
\put(390,57){\small 2}
\put(400,41){$\bar d$}
\put(390,38){\small 3}
\put(399,19){$\bar d$}
\put(384,19){\small 4}
\put(343,86){$u$}
\put(333,82){\small 5}
\put(343,6){$\bar d$}
\put(333,11){\small 6}
\put(348,68){$u$}
\put(342,68){\small 1}
\put(342,51){$u$}
\put(336,59){\small 2}
\put(348,42){$\bar d$}
\put(338,45){\small 3}
\put(348,23){$\bar d$}
\put(341,26){\small 4}
\put(317,65){\footnotesize $1^{uu}$}
\put(317,29){\footnotesize $1^{\bar d \bar d}$}
\put(356,57){\footnotesize $1^{uu}$}
\put(356,37){\footnotesize $1^{\bar d \bar d}$}
\end{picture}

\vskip60pt
\begin{picture}(600,60)
\put(90,48){$+$}
\put(107,48){4}
\put(125,45){\line(1,0){18}}
\put(125,47){\line(1,0){17.5}}
\put(125,49){\line(1,0){17}}
\put(125,51){\line(1,0){17}}
\put(125,53){\line(1,0){17.5}}
\put(125,55){\line(1,0){18}}
\put(155,50){\circle{25}}
\put(144,46){\line(1,1){15}}
\put(147,41){\line(1,1){17}}
\put(152.5,38.5){\line(1,1){14}}
\put(164,68){\circle{16}}
\put(164,32){\circle{16}}
\put(175,50){\circle{16}}
\put(183,50){\vector(1,1){12}}
\put(183,50){\vector(1,-1){12}}
\put(171,70){\vector(3,2){21}}
\put(171,70){\vector(3,-1){23}}
\put(171,30){\vector(3,1){23}}
\put(171,30){\vector(3,-2){21}}
\put(203,60){\circle{16}}
\put(203,40){\circle{16}}
\put(212,60){\vector(3,2){21}}
\put(212,60){\vector(3,-1){23}}
\put(212,40){\vector(3,1){23}}
\put(212,40){\vector(3,-2){21}}
\put(239,71){$u$}
\put(227,77){\small 1}
\put(240,54){$u$}
\put(230,57){\small 2}
\put(240,40){$\bar d$}
\put(230,37){\small 3}
\put(239,19){$\bar d$}
\put(227,16){\small 4}
\put(183,86){$u$}
\put(173,81){\small 5}
\put(183,6){$\bar d$}
\put(173,12){\small 6}
\put(188,68){$u$}
\put(182,68){\small 1}
\put(182,56){\small $u$}
\put(187,48){\small 2}
\put(181,37){\small $\bar d$}
\put(192,44){\small 3}
\put(188,24){$\bar d$}
\put(182,25){\small 4}
\put(157,65){\footnotesize $1^{uu}$}
\put(157,29){\footnotesize $1^{\bar d \bar d}$}
\put(168,47){\footnotesize $1^{u \bar d}$}
\put(196,57){\footnotesize $1^{uu}$}
\put(196,37){\footnotesize $1^{\bar d \bar d}$}
\end{picture}

\newpage


\vskip60pt
\begin{picture}(600,100)
\put(0,45){\line(1,0){18}}
\put(0,47){\line(1,0){17.5}}
\put(0,49){\line(1,0){17}}
\put(0,51){\line(1,0){17}}
\put(0,53){\line(1,0){17.5}}
\put(0,55){\line(1,0){18}}
\put(30,50){\circle{25}}
\put(19,46){\line(1,1){15}}
\put(22,41){\line(1,1){17}}
\put(27.5,38.5){\line(1,1){14}}
\put(40,68){\circle{16}}
\put(40,32){\circle{16}}
\put(51,50){\circle{16}}
\put(48,70){\vector(3,2){19}}
\put(48,70){\vector(3,-1){22}}
\put(59,50){\vector(3,2){14}}
\put(59,50){\vector(3,-2){14}}
\put(48,30){\vector(3,1){22}}
\put(48,30){\vector(3,-2){19}}
\put(70,80){$u$}
\put(60,84){\small 1}
\put(73,63){$u$}
\put(63,67){\small 2}
\put(75,53){$u$}
\put(61,56){\small 3}
\put(75,41){$\bar d$}
\put(61,39){\small 4}
\put(73,30){$\bar d$}
\put(63,27){\small 5}
\put(70,13){$\bar d$}
\put(60,9){\small 6}
\put(33,65){\footnotesize $1^{uu}$}
\put(33,29){\footnotesize $1^{\bar d \bar d}$}
\put(44,47){\footnotesize $1^{u \bar d}$}
\put(90,48){$=$}
\put(110,45){\line(1,0){19}}
\put(110,47){\line(1,0){21}}
\put(110,49){\line(1,0){23}}
\put(110,51){\line(1,0){23}}
\put(110,53){\line(1,0){21}}
\put(110,55){\line(1,0){19}}
\put(133,65){\circle{16}}
\put(142,50){\circle{16}}
\put(133,35){\circle{16}}
\put(141,68){\vector(1,1){13}}
\put(141,68){\vector(3,-1){18}}
\put(150,50){\vector(3,2){16}}
\put(150,50){\vector(3,-2){16}}
\put(141,32){\vector(3,1){18}}
\put(141,32){\vector(1,-1){13}}
\put(158,80){$u$}
\put(143,79){\small 1}
\put(160,65){$u$}
\put(150,67){\small 2}
\put(170,55){$u$}
\put(158,48){\small 3}
\put(170,39){$\bar d$}
\put(163,44){\small 4}
\put(160,28){$\bar d$}
\put(150,27){\small 5}
\put(158,10){$\bar d$}
\put(143,13){\small 6}
\put(126,62){\footnotesize $1^{uu}$}
\put(135,47){\footnotesize $1^{u \bar d}$}
\put(126,32){\footnotesize $1^{\bar d \bar d}$}
\put(183,48){$+$}
\put(198,48){2}
\put(212,45){\line(1,0){18}}
\put(212,47){\line(1,0){17.5}}
\put(212,49){\line(1,0){17}}
\put(212,51){\line(1,0){17}}
\put(212,53){\line(1,0){17.5}}
\put(212,55){\line(1,0){18}}
\put(242,50){\circle{25}}
\put(231,46){\line(1,1){15}}
\put(234,41){\line(1,1){17}}
\put(239.5,38.5){\line(1,1){14}}
\put(242,30){\circle{16}}
\put(242,22){\vector(2,-3){9}}
\put(242,22){\vector(-2,-3){9}}
\put(262,50){\circle{16}}
\put(270,50){\vector(3,2){17}}
\put(270,50){\vector(3,-2){17}}
\put(247,61){\vector(1,0){40}}
\put(247,39){\vector(1,0){40}}
\put(295,60){\circle{16}}
\put(295,40){\circle{16}}
\put(303,61){\vector(3,1){20}}
\put(303,61){\vector(3,-1){20}}
\put(303,39){\vector(3,1){20}}
\put(303,39){\vector(3,-1){20}}
\put(328,70){$u$}
\put(313,70){\small 1}
\put(328,53){$u$}
\put(307,51){\small 2}
\put(328,41){$u$}
\put(313,46){\small 3}
\put(328,24){$\bar d$}
\put(313,24){\small 4}
\put(224,10){$\bar d$}
\put(234,1){\small 5}
\put(255,10){$\bar d$}
\put(245,1){\small 6}
\put(270,66){$u$}
\put(260,65){\small 1}
\put(280,52){\small $u$}
\put(270,54){\small 2}
\put(280,44){\small $u$}
\put(270,41){\small 3}
\put(270,26){$\bar d$}
\put(260,29){\small 4}
\put(235,27){\footnotesize $1^{\bar d \bar d}$}
\put(255,47){\footnotesize $1^{uu}$}
\put(288,57){\footnotesize $1^{uu}$}
\put(288,37){\footnotesize $1^{u \bar d}$}
\put(341,48){$+$}
\put(356,48){2}
\put(370,45){\line(1,0){18}}
\put(370,47){\line(1,0){17.5}}
\put(370,49){\line(1,0){17}}
\put(370,51){\line(1,0){17}}
\put(370,53){\line(1,0){17.5}}
\put(370,55){\line(1,0){18}}
\put(400,50){\circle{25}}
\put(389,46){\line(1,1){15}}
\put(392,41){\line(1,1){17}}
\put(397.5,38.5){\line(1,1){14}}
\put(400,30){\circle{16}}
\put(400,22){\vector(2,-3){9}}
\put(400,22){\vector(-2,-3){9}}
\put(420,50){\circle{16}}
\put(428,50){\vector(3,2){17}}
\put(428,50){\vector(3,-2){17}}
\put(405,61){\vector(1,0){40}}
\put(405,39){\vector(1,0){40}}
\put(453,60){\circle{16}}
\put(453,40){\circle{16}}
\put(461,61){\vector(3,1){20}}
\put(461,61){\vector(3,-1){20}}
\put(461,39){\vector(3,1){20}}
\put(461,39){\vector(3,-1){20}}
\put(486,70){$u$}
\put(471,70){\small 1}
\put(486,53){$u$}
\put(465,51){\small 2}
\put(486,41){$u$}
\put(471,46){\small 3}
\put(486,24){$\bar d$}
\put(471,24){\small 4}
\put(382,10){$\bar d$}
\put(392,1){\small 5}
\put(413,10){$\bar d$}
\put(403,1){\small 6}
\put(428,66){$u$}
\put(418,65){\small 1}
\put(435,50){\small $u$}
\put(428,54){\small 2}
\put(440,44){\small $\bar d$}
\put(428,41){\small 3}
\put(428,29){$u$}
\put(418,29){\small 4}
\put(393,27){\footnotesize $1^{\bar d \bar d}$}
\put(413,47){\footnotesize $1^{u \bar d}$}
\put(446,57){\footnotesize $1^{uu}$}
\put(446,37){\footnotesize $1^{u \bar d}$}
\end{picture}

\vskip60pt
\begin{picture}(600,60)
\put(90,48){$+$}
\put(105,48){2}
\put(119,45){\line(1,0){18}}
\put(119,47){\line(1,0){17.5}}
\put(119,49){\line(1,0){17}}
\put(119,51){\line(1,0){17}}
\put(119,53){\line(1,0){17.5}}
\put(119,55){\line(1,0){18}}
\put(149,50){\circle{25}}
\put(138,46){\line(1,1){15}}
\put(141,41){\line(1,1){17}}
\put(146.5,38.5){\line(1,1){14}}
\put(149,30){\circle{16}}
\put(149,22){\vector(2,-3){9}}
\put(149,22){\vector(-2,-3){9}}
\put(169,50){\circle{16}}
\put(177,50){\vector(3,2){17}}
\put(177,50){\vector(3,-2){17}}
\put(154,61){\vector(1,0){40}}
\put(154,39){\vector(1,0){40}}
\put(202,60){\circle{16}}
\put(202,40){\circle{16}}
\put(210,61){\vector(3,1){20}}
\put(210,61){\vector(3,-1){20}}
\put(210,39){\vector(3,1){20}}
\put(210,39){\vector(3,-1){20}}
\put(235,70){$\bar d$}
\put(220,70){\small 1}
\put(235,53){$\bar d$}
\put(214,51){\small 2}
\put(235,41){$u$}
\put(220,46){\small 3}
\put(235,24){$\bar d$}
\put(220,24){\small 4}
\put(131,10){$u$}
\put(141,1){\small 5}
\put(162,10){$u$}
\put(152,1){\small 6}
\put(177,65){$\bar d$}
\put(167,65){\small 1}
\put(189,49){\small $\bar d$}
\put(177,54){\small 2}
\put(187,44){\small $u$}
\put(177,41){\small 3}
\put(177,26){$\bar d$}
\put(167,29){\small 4}
\put(142,27){\footnotesize $1^{uu}$}
\put(162,47){\footnotesize $1^{u \bar d}$}
\put(195,57){\footnotesize $1^{\bar d \bar d}$}
\put(195,37){\footnotesize $1^{u \bar d}$}
\put(248,48){$+$}
\put(263,48){2}
\put(277,45){\line(1,0){18}}
\put(277,47){\line(1,0){17.5}}
\put(277,49){\line(1,0){17}}
\put(277,51){\line(1,0){17}}
\put(277,53){\line(1,0){17.5}}
\put(277,55){\line(1,0){18}}
\put(307,50){\circle{25}}
\put(296,46){\line(1,1){15}}
\put(299,41){\line(1,1){17}}
\put(304.5,38.5){\line(1,1){14}}
\put(307,30){\circle{16}}
\put(307,22){\vector(2,-3){9}}
\put(307,22){\vector(-2,-3){9}}
\put(327,50){\circle{16}}
\put(335,50){\vector(3,2){17}}
\put(335,50){\vector(3,-2){17}}
\put(312,61){\vector(1,0){40}}
\put(312,39){\vector(1,0){40}}
\put(360,60){\circle{16}}
\put(360,40){\circle{16}}
\put(368,61){\vector(3,1){20}}
\put(368,61){\vector(3,-1){20}}
\put(368,39){\vector(3,1){20}}
\put(368,39){\vector(3,-1){20}}
\put(393,70){$\bar d$}
\put(378,70){\small 1}
\put(393,53){$\bar d$}
\put(372,51){\small 2}
\put(393,41){$u$}
\put(378,46){\small 3}
\put(393,24){$\bar d$}
\put(378,24){\small 4}
\put(289,10){$u$}
\put(299,1){\small 5}
\put(320,10){$u$}
\put(310,1){\small 6}
\put(335,65){$\bar d$}
\put(325,65){\small 1}
\put(343,48){\small $\bar d$}
\put(335,54){\small 2}
\put(347,44){\small $\bar d$}
\put(335,41){\small 3}
\put(335,29){$u$}
\put(325,29){\small 4}
\put(300,27){\footnotesize $1^{uu}$}
\put(320,47){\footnotesize $1^{\bar d \bar d}$}
\put(353,57){\footnotesize $1^{\bar d \bar d}$}
\put(353,37){\footnotesize $1^{u \bar d}$}
\end{picture}

\vskip60pt
\begin{picture}(600,60)
\put(90,48){$+$}
\put(105,48){4}
\put(119,45){\line(1,0){18}}
\put(119,47){\line(1,0){17.5}}
\put(119,49){\line(1,0){17}}
\put(119,51){\line(1,0){17}}
\put(119,53){\line(1,0){17.5}}
\put(119,55){\line(1,0){18}}
\put(149,50){\circle{25}}
\put(138,46){\line(1,1){15}}
\put(141,41){\line(1,1){17}}
\put(146.5,38.5){\line(1,1){14}}
\put(149,30){\circle{16}}
\put(149,22){\vector(2,-3){9}}
\put(149,22){\vector(-2,-3){9}}
\put(169,50){\circle{16}}
\put(177,50){\vector(3,2){17}}
\put(177,50){\vector(3,-2){17}}
\put(154,61){\vector(1,0){40}}
\put(154,39){\vector(1,0){40}}
\put(202,60){\circle{16}}
\put(202,40){\circle{16}}
\put(210,61){\vector(3,1){20}}
\put(210,61){\vector(3,-1){20}}
\put(210,39){\vector(3,1){20}}
\put(210,39){\vector(3,-1){20}}
\put(235,70){$u$}
\put(220,70){\small 1}
\put(235,53){$u$}
\put(214,51){\small 2}
\put(235,41){$\bar d$}
\put(220,46){\small 3}
\put(235,24){$\bar d$}
\put(220,24){\small 4}
\put(131,10){$u$}
\put(141,1){\small 5}
\put(162,10){$\bar d$}
\put(152,1){\small 6}
\put(177,65){$u$}
\put(167,65){\small 1}
\put(185,50){\small $u$}
\put(177,54){\small 2}
\put(189,44){\small $\bar d$}
\put(177,41){\small 3}
\put(177,26){$\bar d$}
\put(167,29){\small 4}
\put(142,27){\footnotesize $1^{u \bar d}$}
\put(162,47){\footnotesize $1^{u \bar d}$}
\put(195,57){\footnotesize $1^{uu}$}
\put(195,37){\footnotesize $1^{\bar d \bar d}$}
\put(248,48){$+$}
\put(263,48){4}
\put(277,45){\line(1,0){18}}
\put(277,47){\line(1,0){17.5}}
\put(277,49){\line(1,0){17}}
\put(277,51){\line(1,0){17}}
\put(277,53){\line(1,0){17.5}}
\put(277,55){\line(1,0){18}}
\put(307,50){\circle{25}}
\put(296,46){\line(1,1){15}}
\put(299,41){\line(1,1){17}}
\put(304.5,38.5){\line(1,1){14}}
\put(304,62){\vector(2,1){40}}
\put(304,38){\vector(2,-1){40}}
\put(324.5,60){\circle{16}}
\put(324.5,40){\circle{16}}
\put(329,67){\vector(1,1){15}}
\put(329,33){\vector(1,-1){15}}
\put(329,67){\vector(1,-1){17}}
\put(329,33){\vector(1,1){17}}
\put(352,84){\circle{16}}
\put(354,50){\circle{16}}
\put(352,16){\circle{16}}
\put(360,86){\vector(3,2){18}}
\put(360,86){\vector(3,-2){18}}
\put(362,50){\vector(3,2){18}}
\put(362,50){\vector(3,-2){18}}
\put(360,14){\vector(3,2){18}}
\put(360,14){\vector(3,-2){18}}
\put(382,95){$u$}
\put(370,100){\small 1}
\put(382,70){$u$}
\put(372,66){\small 2}
\put(382,58){$u$}
\put(368,58){\small 3}
\put(382,33){$\bar d$}
\put(368,35){\small 4}
\put(382,21){$\bar d$}
\put(372,27){\small 5}
\put(382,-4){$\bar d$}
\put(372,-6){\small 6}
\put(327,80){$u$}
\put(317,77){\small 1}
\put(340,69){$u$}
\put(335,66){\small 2}
\put(340,58){$u$}
\put(333,52){\small 3}
\put(340,35){$\bar d$}
\put(333,42){\small 4}
\put(340,23){$\bar d$}
\put(335,29){\small 5}
\put(327,12){$\bar d$}
\put(317,17){\small 6}
\put(317.5,57){\footnotesize $1^{uu}$}
\put(317.5,37){\footnotesize $1^{\bar d \bar d}$}
\put(345,81){\footnotesize $1^{uu}$}
\put(347,47){\footnotesize $1^{u \bar d}$}
\put(345,13){\footnotesize $1^{\bar d \bar d}$}
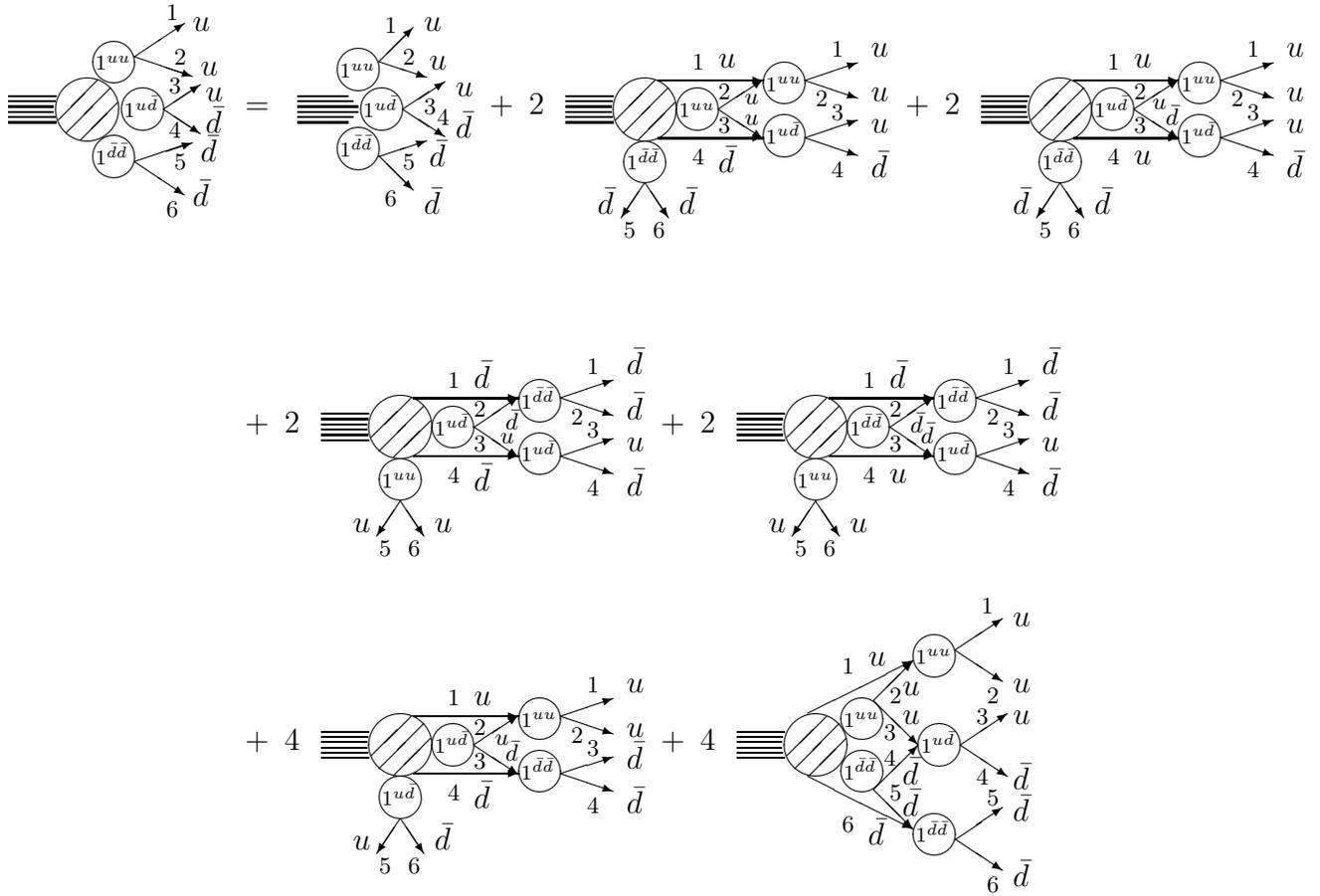
\put(0,-30){Fig. 3. Graphic representation of the equations for the
six-quark subamplitudes $A_l$ $(l=1, 2, 3)$}
\put(0,-50){in the case of baryonium $uuu \bar d \bar d \bar d$ $IJ=33$.}
\end{picture}

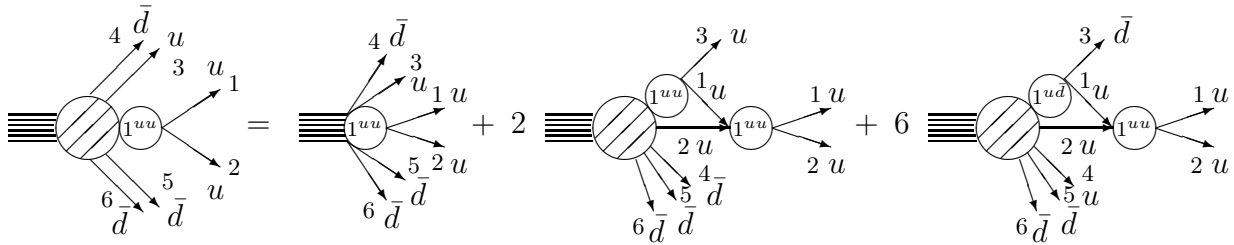
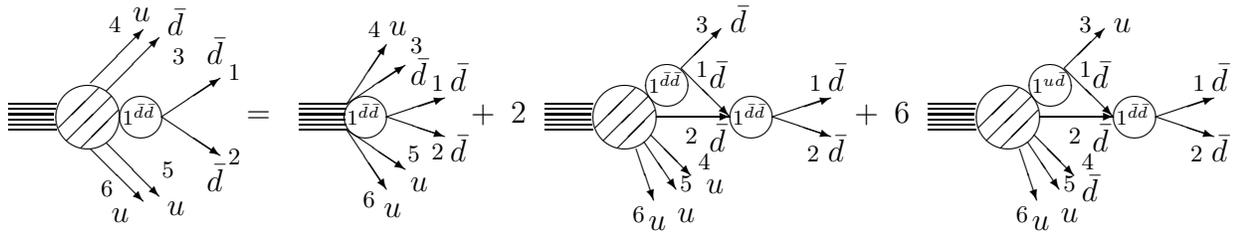
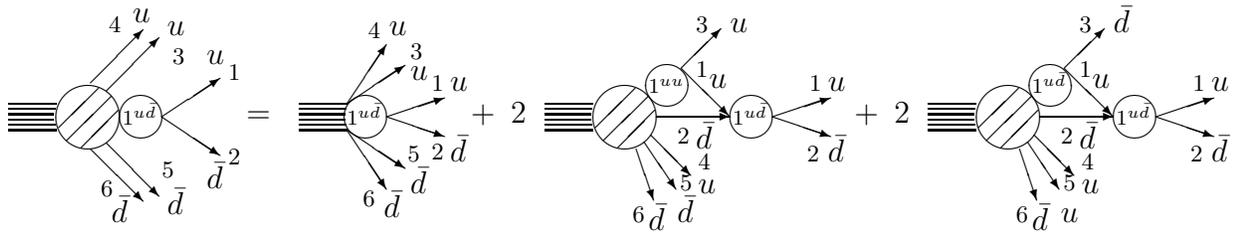
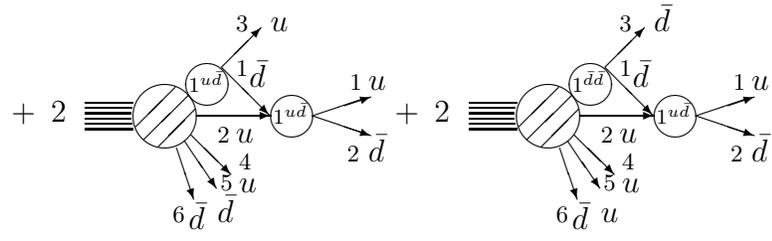
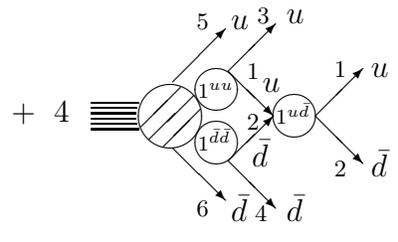
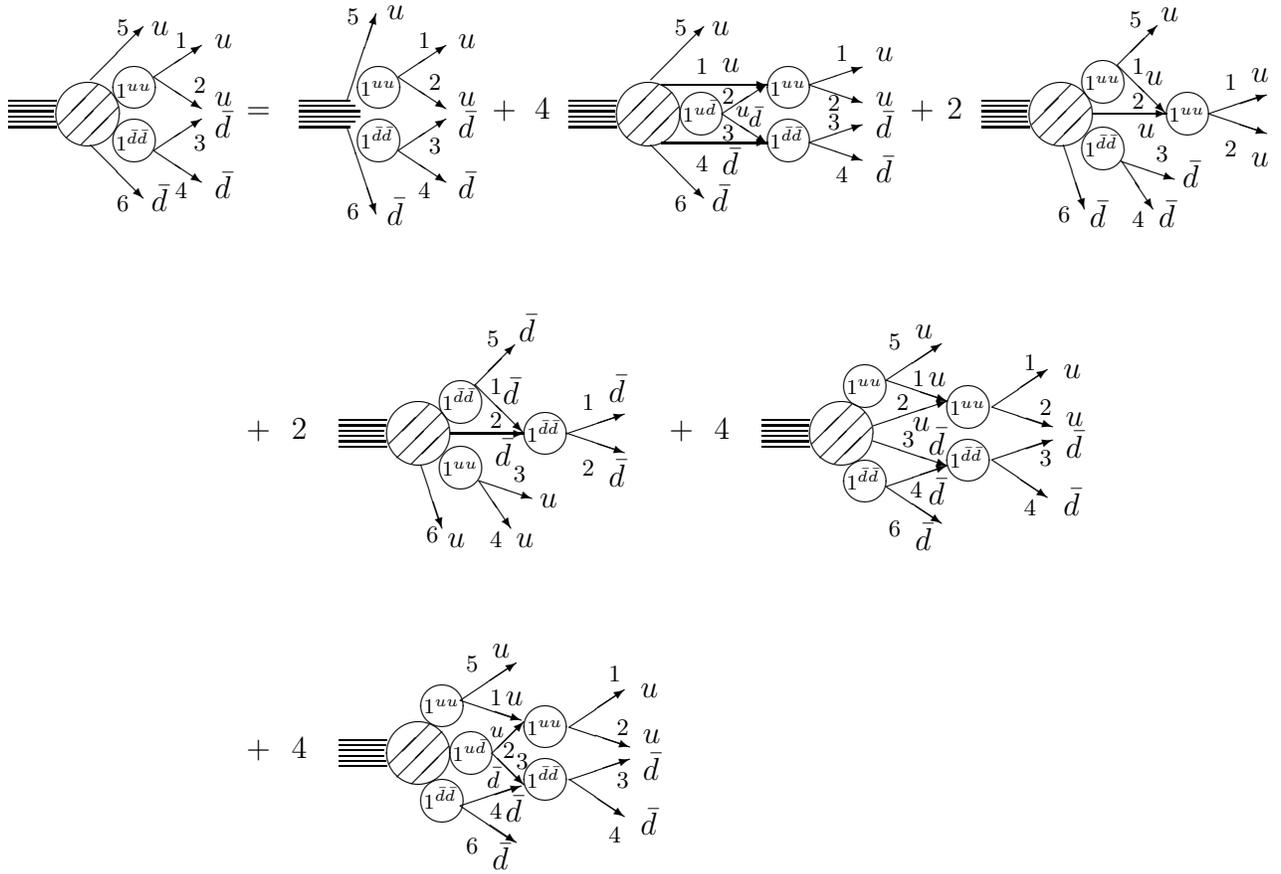
\end{document}